\documentclass[pre,aps,reprint,amsmath]{revtex4-1}

\usepackage[T1]{fontenc}
\usepackage{graphicx}
\usepackage{fancyvrb}
\usepackage{amsmath,amssymb,amstext}
\usepackage{url}
\usepackage{microtype}
\usepackage[utf8]{inputenc}
\usepackage{lmodern}
\usepackage{bbold}
\usepackage{color}
\usepackage[hidelinks]{hyperref}

\begin{document}
\title{ Phase separation of active Brownian particles in two dimensions: \\
Anything for a quiet life}

\author{Sophie Hermann}
\affiliation{Theoretische Physik II, Physikalisches Institut, Universit{\"a}t Bayreuth, D-95447 Bayreuth, Germany}

\author{Daniel de las Heras}
\affiliation{Theoretische Physik II, Physikalisches Institut, Universit{\"a}t Bayreuth, D-95447 Bayreuth, Germany}

\author{Matthias Schmidt}
\affiliation{Theoretische Physik II, Physikalisches Institut, Universit{\"a}t Bayreuth, D-95447 Bayreuth, Germany}
\email{Matthias.Schmidt@uni-bayreuth.de}

\date{27 November 2020,
  Mol.\ Phys. {\it Gerhard Findenegg Memorial Issue},
DOI:\href{https://doi.org/10.1080/00268976.2021.1902585}{10.1080/00268976.2021.1902585}}

\begin{abstract}
Active Brownian particles display self-propelled movement, which can be modelled as arising from a one-body force. 
Although their interparticle interactions are purely repulsive, for strong self propulsion the swimmers phase separate into dilute and dense phases.
We describe in detail a recent theory (\href{https://link.aps.org/doi/10.1103/PhysRevE.100.052604}{Phys. Rev. E \textbf{100}, 052604 (2019)}; \href{https://doi.org/10.1103/PhysRevLett.123.268002}{Phys. Rev. Lett. \textbf{123}, 26802 (2019)}) for such motility induced phase separation. Starting from the continuity equation and the force density balance, the description is based on four superadiabatic contributions to the internal force density. Here the superadiabatic forces are due to the flow in the system and they act on top of the adiabatic forces that arise from the equilibrium free energy. Phase coexistence is described by bulk state functions and agrees quantitatively with Brownian dynamics simulation results from the literature. We describe in detail all analytical steps to fully resolve the spatial and orientational dependence of the one-body density and current. The decomposition into angular Fourier series leads to coupling of total density, polarization and all higher modes. We  describe the power functional approach, including the kinematic dependence of the superadiabatic force fields and the quiet life effect that pushes particles from fast to slow regions, and hence induces the phase separation.
\end{abstract}
\maketitle
\section{Introduction}

Active Brownian particles (ABPs) undergo self-propelled motion and are considered to be a minimal model for the more general class of active matter. Systems of active Brownian particles often are seen as a prototype to develop and test more general approaches to nonequilibrium Statistical Physics, see e.g.\ Refs. \cite{schmidt2019,postension, krinninger2019,krinninger2016}. One particular phenomenon of Brownian swimmers that has attracted much current interest is motility induced phase separation (MIPS) (see e.g. the reviews  \cite{rev1,rev2,rev3,rev4,rev5,rev6,rev7}). 
MIPS occurs for repulsive particles with sufficient strength of their self propulsion. The system separates into a dense and a dilute phase, similar to liquid-vapour phase separation in equilibrium.
The literature contains a range of theoretical descriptions of such systems.  
A broad overview over the current literature can be found in Refs.~\cite{rev1,rev2,rev3,rev4,rev5,rev6,rev7} and also in the introduction of Ref. \cite{krinninger2019},  including different descriptions of swimming such as chemotaxis, hydrodynamics and active Brownian motion, as leading to collective phenomena such as clustering and MIPS.
In the following we summarize several selected theories for MIPS in active Brownian particles. 

Experimentally realized active particles generate hydrodynamic solvent flow, which induces hydrodynamic interactions between the particles. To include these interactions in theoretical models is interesting and can influence MIPS non-trivially \cite{stark2014,gompper2018}. Including hydrodynamic interaction was found to either enhance \cite{stark2014} or suppress \cite{gompper2018} phase separation of spherical swimmers in a narrow slit.
 
Speck and co-workers \cite{bialke2013,menzel2014,menzel2015} developed a model for two-dimensional active disks with repulsive interparticle interactions. These authors neglected hydrodynamical interactions and derived effective hydrodynamic equations by integration of the $N$-body Smoluchowski equation using two different closure relations \cite{bialke2013}. The Smoluchowski equation is the Fokker-Planck equation (as a spatio-temporal partial differential equation for the propability density distribution) for overdamped motion.  Furthermore these authors carried out a linear stability analysis \cite{bialke2013,menzel2014}. Comparison with the scaling behaviour of a small perturbation results in an effective Cahn-Hilliard equation. They translate results of the classical Cahn-Hilliard equation such as the existence of a free energy functional and determine the onset of phase separation from that functional \cite{menzel2014}. 
More detailed considerations as an adiabatic approximation or weakly non-linear analysis were presented in \cite{menzel2015}. These authors compare several phase diagrams from simulation and theory. The phase diagrams coincide for low gaseous densities but deviate for high liquid densities, which was attributed to differing particle compressibilities.

Takatori, Yan and Brady introduced the swim pressure \cite{takatori2014} for two- and three- dimensional active spheres. From this pressure Takatori and Brady construct a mechanical theory, neglecting all hydrodynamic interactions \cite{brady2015}. An active pressure is defined via the combination  of the swim pressure and the interparticle interaction pressure. The corresponding chemical potential follows from a mechanical relation determined from the momentum balance for active particles. The resulting phase diagram of active particles is compared to simulation results. The binodal is defined by the condition of equal values of the chemical potential in the bulk. The spinodal (see Ref. \cite{brady2020} for recent work) is determined via vanishing of the first derivative of the active pressure with respect to the packing fraction. These authors observe that two-dimensional systems phase separate more easily than three-dimensional systems. A detailed study of the influence of dimensionality is given by Stenhammar et~al. \cite{stenhammar2014}. 
Additionally the concepts of thermodynamics were transferred to nonequilibrium \cite{brady2015}. Several different quantities as the Helmholz free energy, the Gibbs free energy, entropy and heat capacity were calculated. But the authors also state that it is not certain whether those relations are applicable.

The mechanical pressure in the two-dimensional system of active Brownian spheres considered by Solon et~al. \cite{solon2015} arises from the particle interaction with the confining walls. A similar (wetting) situation was recently addressed by Neta et~al. \cite{neta2020} in the context of a lattice model. The pressure is calculated using Newton's third law and radial harmonics of zeroth, first and second  order corresponding to total density, polarization and nematic order, respectively. The pressure splits (as does the active pressure in \cite{brady2015}) into an interaction contribution and a swim pressure, which is further decomposed into an ideal and an indirect part. The expression for the pressure turns out to be a state function, which has the same value in coexisting bulk phases and is independent of the wall-particle interactions. However the binodal determined from the mechanical pressure assuming a hard cut-off for the interaction pressure and using the equal area Maxwell construction does not match the coexistence data of active Brownian dynamics simulations.

A disagreement between the Maxwell construction and Brownian dynamics simulations was reported by Paliwal et~al. \cite{utrecht2018}. In contrast to Ref. \cite{solon2015}, the authors constructed from Fokker-Planck calculations an expression for the chemical potential which contains an intrinsic, an external and a swim contribution. The swim contribution arises from a swim potential, which originates from the particle polarization at the interface. 

Active Brownian particles in steady state were considered in several different three-dimensional systems \cite{utrecht2018}: as an ideal gas, as Lennard-Jones swimmers under sedimentation and as Lennard-Jones particles in case of weak phase separation. Additionally two-dimensional particles with Weeks-Chandler-Anderson (WCA) interparticle pair interaction undergoing MIPS were investigated \cite{utrecht2018}.
For active Lennard-Jones swimmers the binodal is found to agree with simulation data. Here the interesting regime is that of low self-propulsion speeds, where activity competes with the entropy vs. energy balance of gas-liquid coexistence in equilibrium. However, for high strength of swimming as is relevant for WCA particles undergoing MIPS the theory clearly overestimates the coexisting densities \cite{utrecht2018}). In simulations the binodal is determined from the value of the orientationally averaged densities in the (gaseous or liquid) portion of a phase separated system. Since the theoretical binodal is derived by the equal area Maxwell construction and therefore using the Gibbs-Duhem relation, it was proposed that this relation is not valid due to high anisotropy in the system. It was concluded that the consideration of interfacial contributions is necessary to determine coexistence densities. This situation is in contrast to equilibrium liquid-vapour phase separation. The approach of \cite{utrecht2018} has some similarities to the work of Takatori and Brady \cite{brady2015} and to that of Solon et~al. \cite{solon2015}, as these authors also introduce pressures and chemical potentials.

Farage et~al. \cite{farage2015} derived an effective interparticle pair interaction potential to describe active particles. Starting from the Langevin equations for active Brownian spheres and averaging with respect to the orientational degrees of freedom leads to an equation of motion with coloured noise. Using the Fox approximation \cite{fox}, which is exact for infinitely fast rotational diffusion, Farage et~al. derive a Fokker-Planck equation. From this relation an effective interparticle interaction potential is determined, which turns out to develop an increasingly negative tail and therefore become more attractive as the particle activity is increased. Here effective potential can be attractive even when the bare particle-particle interaction potential is purely repulsive. 
The occurrence of an effective interaction potential with attraction makes the existence of a phase-separated state at high activity plausible.

The effective interparticle pair interaction potential determined in \cite{farage2015} was used in the context of dynamical density functional theory \cite{rene2016}. 
The authors used the dynamical density functional theory to determine interfacial quantities, such as for the free interface, for wetting and capillary condensation and evaporation of both purely repulsive and Lennard-Jones particles.

Despite the current interest in MIPS and the in simulations well reproducible density distribution across the interface, the number of theoretical descriptions considering the interface between the gas and the liquid is relatively limited.
One approach is to obtain interfacial quantites from simulation results. The authors of Ref. \cite{utrecht2018} derived thereby the position-dependent chemical potential and pressure contributions. Bialké et~al. \cite{negTension2015} derived non vanishing components of the pressure tensor.
Analogous to capillary wave theory the profile of the interface is decomposed in Fourier modes and from that decomposition the average interfacial width is determined \cite{negTension2015, speck2016}. The calculated interfacial surface tension was remarkably reported to be negative \cite{negTension2015}, which was challenged in Ref. \cite{postension} on theoretical grounds.

This result lead to a controversy about the sign of the interfacial tension. 
Brady and his coworkers \cite{bradytension} argue that the negative sign results from the swim pressure, which they describe as not a true pressure. Therefore these authors determine a correction for the pressure tensor, which changes the interfacial tension from a highly negative value to approximately zero. 
Speck \cite{speck2020,speck12020} derived the tension analytically from a van der Waals theory for active discs using both the mechanical and the thermodynamical route. He found both signs, positive and negative, depending on the chosen route.
The generalized thermodynamics by Solon et~al. \cite{solon2018} also supports both positive and negative sign of the interfacial tension and predicts a negative value for active Brownian particles. 
The underlying interfacial theory is based on an effective Cahn-Hilliard equation. These authors determined  the coexisting densities from an explicit violation  of the Maxwell construction. 
A violation of the common equal area construction is further found in Refs. \cite{solon2015, utrecht2018}. It was suggested that the effect might be due to the influence of the interface on the coexisting bulk densities \cite{utrecht2018,solon2018}. 

The formally exact power functional theory (PFT) \cite{pft2013} is based on a splitting of the effective one-body force field that is generated from the interparticle interactions. Both adiabatic and superadiabatic contributions occur. 
The former are an instantaneous density functional.
These contributions are thus independent of the current and similar to what underlies dynamical density functional theory.
The latter, superadiabatic contributions \cite{pft2013,fortini2014} account for nonequilibrium effects that go beyond the adiabatic approximation. Hence superadiabatic terms are specific to the form of the considered nonequilibrium situation and in particular they depend on the flow that occurs in the system. 
For ABPs a range of physically distinct force types was identified based on analysis of computer simulation data \cite{krinninger2016,schmidt2019,krinninger2019}, including drag \cite{krinninger2016,schmidt2019} and forces that arise from superadiabatic pressure and chemical potential contributions \cite{schmidt2019}. The general PFT framework \cite{pft2013} admits straightforward formulation for ABPs \cite{krinninger2019}. Crucially though, the formally defined superadiabatic forces were modelled and made very concrete in \cite{schmidt2019} using both correlator expressions that are suitable to sampling in Brownian dynamics simulations, and semi-local kinematic functional forms, i.e. expressions that express the respective superadiabatic force depending on the (orientation-resolved) density and velocity profiles. Using a small number of adjustable parameters that determine the strength of each superadiabatic force contribution, and which were obtained from comparison to the simulation data \cite{schmidt2019}, the otherwise standalone theory predicts very satisfactory agreement for both the MIPS binodal and spinodal against simulation results \cite{utrecht2018,stenhammar2014}. Crucially, the curiously high density of the gas at coexistence, even under strong driving where coexistence is very broad in density, is captured correctly. The spinodal on the gas side shows similar behaviour and again the agreement with simulation results for the spinodal is excellent, cf. Fig. 2 in \cite{schmidt2019}. As the theory does not discriminate between a mechanical and a thermodynamic point of view, but rather unifies both consistently, all further quantities are unanimous. In particular there is a unique interfacial tension associated with the free interface in MIPS. Based on a semi-local superadiabatic approximation the value of the interfacial tension is found to be positive. The interfacial treament is consistent with the exact result of the interfacial polarization being a statefunction of the adjacent bulk phases \cite{sumrule}, as verified theoretically \cite{auschra2020} and experimentally \cite{soeker2020} for a single laser-nudged microswimmer in a dedicated force-free particle trap.

PFT ascertains that the superadiabatic forces are specific to the type of interparticle interaction potential. The functional dependence of these forces is on density and flow, but they are independent of the external driving (swim force). This formal result has been validated by the universality of the occurring superadiabatic functionals (for repulsive particles). Examples include lane formation in counterdriven binary mixtures \cite{geigenfeind,dzubiella2002,dzubiella2003,dzubiella2004}, memory-induced motion reversal in Brownian liquids \cite{treffenstaedt}, shear-induced deconfinement of hard disks \cite{jahreis2020}, and flow \cite{prl2020, velocitygradient,stuhlmueller2018} and custom flow \cite{renner2019} in Brownian dynamics.
In the present contribution we present in detail the underlying derivations of the theory developed in \cite{schmidt2019}, which describes both the bulk and the interfacial behaviour of this active system.

The theory of \cite{schmidt2019} for ABPs involves rotational degrees of freedom and it is complex both in concept and in execution. Ref. \cite{schmidt2019} only reported the key results and does neither give detailed derivations nor presents the inherent theoretical structure at the fore. Rather, care is taken to back up each step against (position- and orientation-resolved) computer simulation data. Here we  supply all theoretical background, describe the theory based on stand-alone reasoning, and extend and complete the description of Ref.  \cite{schmidt2019} in order to aid future applications of our framework.

This paper is organized as follows. 
In Section \ref{chap:problem} we describe, based on the continuity equation and the force density balance, phase coexistence of active Brownian disks in steady state. Four different superadiabatic (above adiabatic) force density distributions are introduced (Sec. \ref{chap:Fint}) and defined.
Here the superadiabatic force density $\textbf{F}_\text{sup,0}$ is a spherical drag term. In bulk it acts against the particle orientation $\boldsymbol{\omega}$ which indicates the direction of self-propulsion and is associated with a force density. A non-spherical correction to the drag term is given by the force density $\textbf{F}_\text{sup,1}$, an interfacial drag term. It points along $\boldsymbol{\omega}^*$ which is $\boldsymbol{\omega}$ mirrored with respect to the interface normal. The superadiabatic force density $\textbf{F}_\text{sup,2}$ corresponds to a superadiabatic pressure.
The so-called quiet life chemical potential term generates the force density $\textbf{F}_\text{sup,3}$ and opposes (approximately) the adiabatic force density $\textbf{F}_\text{ad}$, where the latter acts perpendicular to the interface towards the active gas.  
Using these superadiabatic forces and a Fourier decomposition one determines recursive relations (cf. Sec. \ref{chap:recursion}) and approximative expressions (cf. Sec.  \ref{chap:rho}) for the one-body density and current. 

We introduce the superadiabatic force correlators (cf. Sec. \ref{chap:Fsup}) and the swim correlator expressions (cf. Sec. \ref{chap:Fext}).
This allows us to determine that the resulting swim pressure is cancelled via a superadiabatic pressure. Hence both pressures have no influence on the coexistence densities. 
 Finally, we show the derivation of the phase diagram via a Maxwell equal area construction (see Sec. \ref{chap:phase}). We conclude and give a short outlook in Sec. \ref{chap:conclusions}.

In four appendices we provide additional analytical details. We describe the bulk fluid limit of ABPs (Appendix \ref{chap:bulk}). We explicitly show the derivation for the $y$-component of the current (Appendix~\ref{chap:y}) and formally introduce two additional superadiabatic force densities, $\textbf{F}_\text{sup,4}$ and $\textbf{F}_\text{sup,5}$, to improve the treatment of the ideal contribution (Appendix~\ref{chap:Fsup45}). The derivation of the MIPS critical point is also shown (Appendix~\ref{chap:crit}).

\section{Motility-induced phase separation} 
\label{chap:mips}
\subsection{Formulation of the problem}
\label{chap:problem}

Consider a (two dimensional) suspension of active Brownian disks with constant free swim speed $s$ or equivalently constant self-propulsion force along their particle orientation $\boldsymbol{\omega}$. The swimmers undergo both translational diffusion and rotational diffusion with corresponding diffusion constants $D$ and $D_\text{rot}$. We neglect hydrodynamical interactions. 
The interactions between the particles are assumed to be purely repulsive and modelled by e.g.\ the Weeks-Chandler-Anderson pair potential \cite{schmidt2019}. 

We aim at constructing a theoretical description for this physical system, based in the one-body level of dynamic correlation functions. 
The starting point of our considerations is the continuity equation for the one-body density $\rho(\textbf{r},\boldsymbol{\omega})$ and the one-body current $\textbf{J}(\textbf{r},\boldsymbol{\omega})$, given by
\begin{align}
\frac{\partial \rho}{\partial t}  = - \nabla \cdot \textbf{J} -\nabla^{\boldsymbol{\omega}} \cdot \textbf{J}^{\boldsymbol{\omega}}, \label{eq:continuity}
\end{align}
where $\nabla$ indicates a spatial and $\nabla^{\boldsymbol{\omega}}$ an orientational derivative. In case of steady state, as considered in the following, the temporal change in density vanishes, i.e. $\partial\rho / \partial t = 0$. Since there are no external torques acting on the spheres (in contrast to active spinners \cite{vitelli,loewen,loewen2}), the rotational current $\textbf{J}^{\boldsymbol{\omega}}$ is purely diffusive and can be expressed as
\begin{align}
\textbf{J}^{\boldsymbol{\omega}} = - D_{\text{rot}} \nabla^{\boldsymbol{\omega}} \rho. \label{eq:Jw}
\end{align}

To simplify the description, the interface between both phases is assumed to be parallel to the $y$-axis, so the system is translational invariant along this direction.
Since we consider a two-dimensional system, any dependence on position $\textbf{r}$ reduces to an $x$-dependence and dependence on particle orientation $\boldsymbol{\omega} = (\cos\varphi, \sin\varphi)$ reduces to a $\varphi$-dependence, where $\varphi$ is the angle measured against the $x$-axis. Thus the derivatives simplify in the spacial case to $\partial/\partial x$ and in the angular case to $\partial/\partial \varphi$.
For the theoretical description we focus on phase separated nonequilibrium steady states. We consider cases where the density distribution increases with increasing $x$-values.  
The origin of the $x$-axis is set to the location $\tilde{x}$ of the Gibbs dividing surface. 
It is defined by the value of $\tilde{x}$ which satisfies
\begin{align}
&\int\limits_{-\infty}^{\tilde{x}} \mathrm{d} x \int\limits_{0}^{2\pi} \mathrm{d} \varphi \left( \rho(x,\varphi) - \rho_\text{g} \right)) + \int\limits^{\infty}_{\tilde{x}} \mathrm{d} x \int\limits_{0}^{2\pi} \mathrm{d} \varphi \left( \rho(x,\varphi) - \rho_\text{l} \right)), \nonumber \\
&=0  \label{eq:gibbs}
\end{align}
where $2 \pi \rho_\text{g}$ ($2 \pi \rho_\text{l}$) denotes the coexisting particle number density per unit area in the dilute (dense) bulk phase of the system.
 
In order to deal with the rotational degree of freedom, it is common to integrate the density with respect to the particle orientation, which yields $2 \pi \rho_0(x)$ \cite{stenhammar2013} and additionally take the next higher order term, the polarization $\pi \rho_1(x)$, into account \cite{bialke2013,menzel2014,utrecht2018}, where $\rho_0$ and $\rho_1$ indicate the zeroth and first order Fourier coefficient of the density.  
Using Fourier space representation is beneficial for construction of analytical solutions, see e.g.\ Ref. \cite{sedimentation} for a study of active ideal sedimentation.
To consider the full angular resolved problem, we Fourier decompose the density and the $x$- and $y$-component of the current \cite{schmidt2019}:
\begin{align}
\rho(x,\varphi) = &\sum\limits_{n=0}^{\infty} \rho_n(x) \cos(n \varphi), \label{eq:FTrho} \\
J^x(x, \varphi) = &\sum\limits_{n=1}^{\infty} J^x_n(x) \cos(n \varphi), \label{eq:FTJx}\\
J^y(x, \varphi) = &\sum\limits_{n=1}^{\infty} J^y_n(x) \sin(n \varphi), \label{eq:FTJy}
\end{align}
where $\rho_n$, $J^x_n$ and $J^y_n$ denote the Fourier coefficients of the density and the current, respectively. Referring to densities implies here units of particle number per two-dimensional system volume and radiant, whereas orientationally integrated densities, as e.g. $2 \pi \rho_0$, represent number of particles per volume.

The system is symmetric with respect to an inversion of the $y$-axis and reversal of the sign of angle $\varphi$, i.e. $y \to -y$ and $\varphi \to - \varphi$. This transformation does neither affect the density nor the $x$-component of the current. Here these are even functions with respect to $\varphi$ and all (odd) sine contributions vanish in the Fourier series. However the $y$-component of the current changes sign under this transformation and it is therefore odd with respect to the angle $\varphi$. So the coefficients corresponding to even cosine terms vanish and the series \eqref{eq:FTJy} only contains the sine summands. 

These symmetries could also be observed in the simulation results where e.g.\ the calculated even Fourier coefficients of the $x$-component of the current are found to be about one order of magnitude larger than the odd coefficients \cite{philipDis}, which validates \eqref{eq:FTJx}. Analogous effects hold for the $y$-component of the current and for the density coefficients.
Furthermore, as is common, the zeroth order ($n=0$) of the current has been neglected for simplicity. Therefore the orientationally averaged flux vanishes, $\int \textbf{J} \; \mathrm{d} \varphi = (J^x_0,J^y_0) = \textbf{0}$.

Considering  all simplifications due to symmetries of the system and inserting the expression for the rotational current $\textbf{J}^{\boldsymbol{\omega}}$ \eqref{eq:Jw} reduces the continuity equation \eqref{eq:continuity} to
\begin{align}
\frac{\partial J^x}{\partial x}  = D_\text{rot} \frac{\partial^2 \rho}{\partial \varphi^2}.
\end{align}
Additionally insertion of the Fourier decomposition and evaluation of the second derivative with respect to $\varphi$ yields
\begin{align}
\sum \limits_{n=1}^\infty \frac{\partial J_n^x}{\partial x} \cos( n \varphi)= - \sum \limits_{n=1}^\infty n^2 D_\text{rot} \rho_n \cos( n \varphi).
\end{align}
Separating this expression into orders of $\cos (n \varphi) $ and reordering the terms leads to
\begin{align}
 \rho_n = -\frac{1}{n^2 D_\text{rot}} \frac{\partial}{\partial x} J^x_n \label{eq:rho}
\end{align}
for all $n \geq 1$. 

The translational current (components) can be in principle determined from the force density balance
\begin{align}
\gamma \textbf{J} = \gamma s \boldsymbol{\omega} \rho + \textbf{F}_\text{int} - k_\text{B} T \nabla \rho, \label{eq:balance}
\end{align}
where $\gamma$ denotes the friction constant, $k_\text{B}$ is the Boltzmann constant and $T$ indicates the temperature. The friction force density on the left hand side is balanced by the self-propulsion (first term), the internal force density $\textbf{F}_\text{int}$ (second term) and the thermal diffusion (third term). 
The internal force density is a complex, usually unknown quantity. In such cases the current cannot be calculated from Eq. \eqref{eq:balance} without approximations for $\textbf{F}_\text{int}$. 

A derivation of Eq. \eqref{eq:balance} and the continuity equation \eqref{eq:continuity} is e.g.\ given in Appendix A of Ref. \cite{schmidt2019}.

\subsection{Superadiabatic and adiabatic force densities}
\label{chap:Fint}

To further specify the expression of the internal force density $\textbf{F}_\text{int}$ in Eq. \eqref{eq:balance}, we split it into an adiabatic part, $\textbf{F}_\text{ad}$, and a superadiabatic part, $\textbf{F}_\text{sup}$ \cite{pft2013,fortini2014},
\begin{align} 
 \textbf{F}_\text{int}= \textbf{F}_\text{ad} + \textbf{F}_\text{sup}.
\end{align}
The superadiabatic contribution is further decomposed into four summands \cite{schmidt2019},
\begin{align}
\textbf{F}_\text{sup} = \textbf{F}_\text{sup,0}+\textbf{F}_\text{sup,1}+\textbf{F}_\text{sup,2}+\textbf{F}_\text{sup,3}, \label{eq:Fsup}
\end{align}
so that one obtains for the internal force density the result
\begin{align}
 \textbf{F}_\text{int}= \textbf{F}_\text{ad} + \textbf{F}_\text{sup,0}+  \textbf{F}_\text{sup,1}+  \textbf{F}_\text{sup,2}+  \textbf{F}_\text{sup,3}. \label{eq:Fint}
\end{align}
In the following all five adiabatic and superadiabatic contributions will be introduced and described in detail. 
The adiabatic contribution can be expressed by the relation from density functional theory \cite{evans1979} 
\begin{align}
\textbf{F}_\text{ad} = - \rho \nabla \frac{\delta F_\text{exc}[\rho]}{\delta \rho} \label{eq:Fad}
\end{align}
with the excess (over ideal gas) intrinsic Helmholtz free energy functional $F_\text{exc}[\rho]$. To apply the concept and Eq. \eqref{eq:Fad} to the situation under investigation one has to consider the so-called adiabatic state. This reference state is the corresponding theoretical equilibrium system that has the same density and the same interparticle interactions as the actual system \cite{pft2013,fortini2014}. 

In the present system the swimmers are spherical and only their direction of self-propulsion is indicated by the orientation. As a consequence, the adiabatic state and thereby the excess free energy functional are independent of orientation.
Furthermore there is no exact explicit expression for $F_\text{exc}[\rho]$ known as is not uncommon in density functional theory, but there exists a range of different approximations \cite{evans2016}. 
We will use the local density approximation on the basis of the scaled particle theory equation of state to specify the adiabatic force density \eqref{eq:Fad} further in Sec. \ref{chap:phase}.

Beyond the adiabatic part the remaining contribution to the total internal force density \eqref{eq:Fint} is formed by the superadiabatic force densities. For each of these superadiabatic terms an approximative analytical expression, as well as a physical interpretation will be given in the following. 
The first two terms, $\textbf{F}_\text{sup,0}$ and $\textbf{F}_\text{sup,1}$, express drag contributions. The drag force density $\textbf{F}_\text{sup,0}$ is proportional to the negative flux $\textbf{J}$, hence it directly opposes the motion and is spherical with respect to $\textbf{J}$. The complete dependence on orientation $\boldsymbol{\omega}$ is contained in the dependence of $\textbf{J}$ on $\boldsymbol{\omega}$. The current and the zeroth superadiabatic force density have thereby the same orientational dependence. 
Our explicit approximation of $\textbf{F}_\text{sup,0}$ is 
the orientationally averaged square gradient expansion of the bulk drag force density, a generalization of the bulk drag contribution in Ref. \cite{krinninger2016} and is given as 
\begin{align}
\textbf{F}_\text{sup,0} = - \gamma \frac{\rho_0}{\rho_\text{jam} - \rho_0} [1 + \xi (\nabla \rho_0)^2] \textbf{J}, \label{eq:Fsup0}
\end{align}
where the parameter $\xi>0$ sets the amplitude of the square gradient term and $\rho_\text{jam}=\text{const}$ is the jamming density. 
If the system density $\rho_0$ reaches locally such a high value, then the prefactor in $\textbf{F}_\text{sup,0}$ diverges and no more motion or flow is possible in this region. Therefore the jamming density models in a very simple way the phenomenon of ''dynamical arrest``. Freezing itself is clearly a more complex transition which implies breaking of translational invariance \cite{pagona2018}. 
Ref. \cite{schmidt2019} contains the description of the adjustment of all occurring free parameters in the theory, such as $\rho_\text{jam}$ and $\xi$, to match theoretical results with Brownian dynamics simulation data. 
The bulk drag force density on which the approximation for $\textbf{F}_\text{sup,0}$ is based is given in \cite{krinninger2016} and reproduced in Eq. \eqref{eq:9}, Appendix \ref{chap:bulk}, where the bulk quantities of the system are summarized. 

The second drag force density $\textbf{F}_\text{sup,1}$ is an interfacial correction to $\textbf{F}_\text{sup,0}$ that takes nonspherical drag contributions into account. Our approximative expression is \cite{schmidt2019} 
\begin{align}
\textbf{F}_\text{sup,1} = - s \gamma \frac{\rho_1}{4}\boldsymbol{\omega}^*, \label{eq:Fsup1}
\end{align}
where $\boldsymbol{\omega}^* = (\cos \varphi, - \sin \varphi)$ is the unit vector that denotes the direction against which the interfacial drag acts. This direction is the orientation $\boldsymbol{\omega}$ mirrored with respect to the $x$-axis. Hence $\boldsymbol{\omega}^*$ can be viewed as the complex conjugate of $\boldsymbol{\omega}$, if the $xy$-coordinates are considered to form the complex plane.
As an aside, note that a description of the reversed succession of the two phases, i.e.\ liquid-gas, with increasing value of $x$ requires the introduction of a minus sign (only) in $\textbf{F}_\text{sup,1}$ as given by \eqref{eq:Fsup1}. This change of sign in $\textbf{F}_\text{sup,1}$ is required to describe the simulation data \cite{schmidt2019}.  

One can prove \cite{pft2013,renner2019} that it is possible to express the total internal force density $\textbf{F}_\text{int}$ and hence the expression for $\textbf{F}_\text{sup,1}$ as a kinematic functional, i.e. as an object with a functional dependence of the density and the current, $\textbf{F}_\text{int}([\rho, \textbf{J},\textbf{J}^{\boldsymbol{\omega}}], \textbf{r}, \boldsymbol{\omega},t)$. This might seem entirely reasonable, if one solves the force density balance \eqref{eq:balance} with respect to the self-propulsion contribution
\begin{align}
s \gamma \rho \boldsymbol{\omega} =  \gamma \textbf{J} - \textbf{F}_\text{int} + k_\text{B}T \nabla \rho. \label{eq:balance2}
\end{align}
Here on the right hand side there is no dependence, neither explicit nor hidden, of the swim force density (or of any other external force densities). Thus there has to be a possibility to rewrite $\textbf{F}_\text{sup,1}$ \eqref{eq:Fsup1} as a kinematic expression without usage of the swim speed $s$, which we will treat later. 

The third superadiabatic force density $\textbf{F}_\text{sup,3}$ is determined by the requirement to balance the thermal diffusion and the adiabatic term,
\begin{align}
\textbf{F}_\text{sup,3} - k_\text{B} T \nabla \rho +\textbf{F}_\text{ad}= 0. \label{eq:Fsup3}
\end{align}
This equation will be considered in more detail in Sec. \ref{chap:phase}, where the mechanism of cancellation is described.
Furthermore the internal force density \eqref{eq:Fint} can be expressed as 
\begin{align}
 \textbf{F}_\text{int}= \textbf{F}_\text{sup,0}+  \textbf{F}_\text{sup,1}+  \textbf{F}_\text{sup,2}+ k_\text{B}T \nabla \rho \label{eq:Fint2}
\end{align}
using Eq. \eqref{eq:Fsup3}.

The remaining uniquely identified contribution to the superadiabatic force density after having split off $\textbf{F}_\text{sup,0}$, $\textbf{F}_\text{sup,1}$ and $\textbf{F}_\text{sup,3}$ is contained in $\textbf{F}_\text{sup,2}$. 
Hence this term is determined by the force density equation \eqref{eq:balance}. We require that the second superadiabatic contribution is independent of orientation for simplicity, although a dependence on $\varphi$ is not excluded on principle grounds.

\subsection{Construction of a density recursion formula}
\label{chap:recursion}

An expression for the $x$-component of the $n$th current Fourier coefficient $J_n^x$ follows from insertion of the Fourier decomposition \eqref{eq:FTrho}-\eqref{eq:FTJy} and of the internal force density $\textbf{F}_\text{int}$ \eqref{eq:Fint} into the $x$-component of the force density equation. In detail, as a first step the internal force density $F^x_\text{int}$ in the $x$-component of the force density balance $\eqref{eq:balance}$ is replaced via Eq. \eqref{eq:Fint2} 
 and the kinematic functionals for the zeroth \eqref{eq:Fsup0} and first \eqref{eq:Fsup1} superadiabatic force densities are inserted. The resulting equation is
\begin{align}
\gamma J^x = \; &\gamma s \rho \cos \varphi - \gamma \frac{\rho_0}{\rho_\text{jam} - \rho_0} \left[ 1 + \xi (\nabla \rho_0)^2 \right] J^x \nonumber\\
&- \gamma s \frac{\rho_1}{4} \cos\varphi + F_\text{sup,2}^x. \label{eq:FTbalanceX1}
\end{align}
Using the Fourier decomposition for $\rho$ \eqref{eq:FTrho} and $J^x$ \eqref{eq:FTJx} leads to
\begin{align}
&\gamma \sum\limits_{n=1}^\infty J_n^x \cos(n \varphi) \label{eq:FTbalanceX}\\
&= \; \gamma s \sum\limits_{n=0}^\infty \rho_n \cos(n \varphi) \cos \varphi - \gamma s \frac{\rho_1}{4} \cos\varphi + F_\text{sup,2}^x \nonumber \\
&{\color{white}= }- \gamma \frac{\rho_0}{\rho_\text{jam} - \rho_0} \left[ 1 + \xi (\nabla \rho_0)^2 \right] \sum\limits_{n=1}^\infty J_n^x \cos(n \varphi) \nonumber 
\end{align} 
In order to separate in orders of $\cos(n \varphi)$, the first term on the right hand side of Eq. \eqref{eq:FTbalanceX}, i.e.\ the swim contribution, is rewritten using the trigonometric relation $2 \cos(n \varphi) \cos \varphi = \cos((n-1)\varphi) + \cos((n+1)\varphi)$ as
\begin{align}
&\gamma s \sum\limits_{n=0}^\infty \rho_n \cos(n \varphi) \cos \varphi  \label{eq:cos}\\
&= \frac{\gamma s }{2} \left[ \sum\limits_{n=-1}^\infty \rho_{n+1} \cos(n \varphi) +  \sum\limits_{n=1}^\infty \rho_{n-1} \cos(n \varphi) \right] \nonumber \\
&=\frac{\gamma s}{2}  \left[ \rho_1 + \rho_0 \cos \varphi +  \sum\limits_{n=1}^\infty \left(\rho_{n-1} + \rho_{n+1}\right) \cos(n \varphi)\right]. \nonumber
\end{align}
As a consequence the force density balance \eqref{eq:FTbalanceX} can be reordered as
\begin{align}
F_\text{sup,2}^x = - \gamma s \frac{\rho_1}{2} + &\gamma s \left( \frac{\rho_1}{4} - \frac{\rho_0}{2} \right) \cos \varphi  \nonumber \\ 
+ \gamma \sum\limits_{n=1}^\infty \bigg\{ &J_n^x \left[ 1 + \frac{\rho_0 [1 + \xi (\nabla \rho_0)^2]}{\rho_\text{jam} - \rho_0} \right] \nonumber\\
&- \frac{s}{2} \left( \rho_{n-1} + \rho_{n+1} \right) \bigg\} \cos(n \varphi). \label{eq:FTbalanceX2}
\end{align}
To simplify the relation we define the expression in square brackets in \eqref{eq:FTbalanceX2} as $s/v_\text{f}$, or equivalently
\begin{align}
v_\text{f} = s \frac{1 - \rho_0/\rho_{\text{jam}}}{1+\xi (\nabla \rho_0)^2 \rho_0/\rho_{\text{jam}}}, \label{eq:vf}
\end{align}
which has units of velocity. We refer to $v_\text{f}$ as the forward speed, in which ''forward`` corresponds to the particle orientation and swim direction $\boldsymbol{\omega}$. One can interpret $v_\text{f}$ as a type of orientationally averaged projection of a velocity on the particle direction $\boldsymbol{\omega}$, although it is not proportional to $\int \mathrm{d} \varphi \; \textbf{v} \cdot \boldsymbol{\omega}$. An explicit correlator expression of the forward speed is given below in Eq. \eqref{eq:corrvf}. 
Note that extrapolation of the forward speed $v_\text{f}$ to the limit $v_\text{f}\rightarrow 0$ can be used to determine the value of the jamming density $\rho_\text{jam}$ from simulation data \cite{schmidt2019}.

To satisfy Eq. \eqref{eq:FTbalanceX2} all prefactors of  $\cos(n \varphi)$ have to identically vanish for each value of $n$ separately. This leads to expressions for the current components $J_n^x$ for $n \geq 1$.
In case of $n>1$ one obtains
\begin{align}
J^x_n  = \frac{v_\text{f}}{2}  \left(\rho_{n-1} + \rho_{n+1}\right), \label{eq:Jxn}
\end{align}
by simply setting the expression in curly brackets in \eqref{eq:FTbalanceX2} equal to zero.
For $n=1$ an additional term, composed of two contributions, is proportional to $\cos \varphi$. The first contribution is the $x$-component of the interfacial drag force density $F^x_\text{sup,1}$ and the second contribution originates from the limiting term $n=-1$ in Eq. \eqref{eq:cos}. 
This yields for the first Fourier coefficient of the current the result
\begin{align}
J^x_1 = \frac{v_\text{f}}{2} \left(\rho_{0} +\rho_{2}\right) + \frac{v_\text{f}}{2} \left( \rho_{0} -\frac{\rho_{1}}{2}  \right) = v_\text{f} \left(\rho_{0} - \frac{\rho_{1}}{4} + \frac{\rho_{2}}{2} \right),\label{eq:Jx1}
\end{align}
where $v_\text{f}$ is still given by \eqref{eq:vf}. Finally considering all contributions that are independent of the particle orientation $ \boldsymbol{\omega}$ determines the expression for the second superadiabatic force density
\begin{align}
F_\text{sup,2}^x &= - \gamma s \frac{\rho_1}{2} = \frac{\gamma s}{2D_\text{rot}}  \frac{\partial}{\partial x} J_1 \nonumber\\
&= \frac{\gamma s}{2D_\text{rot}} \frac{\partial}{\partial x} \left[ v_\text{f} \left( \rho_0 - \frac{\rho_1}{4} + \frac{\rho_2}{2} \right) \right], \label{eq:Fsup2x}
\end{align}
which is by definition rotationally invariant, i.e.\ independent of $\varphi$.
Here we have in the first step replaced the first Fourier coefficient of the density $\rho_1$ using Eq. \eqref{eq:rho} and have subsequently applied relation \eqref{eq:Jx1}.

Performing the analogous calculation for the $y$-component of the force density equation \eqref{eq:balance} gives expressions for the current coefficients $J_n^y$ for $n>1$ and $n=1$, which are
\begin{align}
J^y_n  &= \frac{v_\text{f}}{2}  \left(\rho_{n-1} - \rho_{n+1}\right), \label{eq:Jyn}\\
J^y_1 &= v_\text{f} \left(\rho_{0} +\frac{\rho_{1}}{4} - \frac{\rho_{2}}{2} \right). \label{eq:Jy1}
\end{align}
Furthermore it turns out that the $y$-component of the superadiabatic force density vanishes, $F_\text{sup,2}^y = 0$. The derivation of these results is shown in full detail in Appendix~\ref{chap:y}.

Combination of the Fourier decomposed continuity equation \eqref{eq:rho} and the translational current components $J_n^x$ \eqref{eq:Jxn}, \eqref{eq:Jx1} gives the implicit recursive relation, 
\begin{align}
\rho_n &= - \frac{1}{n^2 D_\text{rot}} \frac{\partial}{\partial x} \left[ \frac{v_\text{f}}{2}  \left(\rho_{n-1} + \rho_{n+1}\right) \right], \label{eq:rhon}\\
\rho_1 &= - \frac{1}{ D_\text{rot}} \frac{\partial}{\partial x} \left[ v_\text{f}  \left(\rho_{0} -\frac{\rho_1}{4} + \frac{\rho_{2}}{2}\right) \right], \label{eq:rho1}
\end{align}
for all density coefficients $\rho_n$ with $n\geq 1$. 
For the zeroth density coefficient $\rho_0$ such a relation does not exist since Eq. \eqref{eq:rho} derived from the continuity equation \eqref{eq:continuity} is only valid for $n\geq1$. This is due to the isotropy of the density coefficients $\rho_0$ (recall that this is orientationally independent per definition), which leads to vanishing derivative with respect to $\varphi$.

\subsection{Calculation of one-body density and current}
\label{chap:rho}

By solving the system of coupled differential equations \eqref{eq:rhon} and \eqref{eq:rho1} one can in principle determine all $\rho_n$ and the corresponding current components $J_n^x$ using \eqref{eq:Jxn} or \eqref{eq:Jx1}. Although this recursion relation might appear to be simple at first glance as it only couples two ''neighbouring`` orders $n\pm1$, it does in fact constitute a difficult problem due to the complex dependence of $v_\text{f}$ on $x$. 
 A comparison of the density coefficients $\rho_n$ determined by the Brownian dynamics simulations shows that the magnitude of $\rho_n$ decreases with increasing order $n$ \cite{philipDis}.
It is therefore a reasonable approximation to neglect higher order density coefficients in the recursion relation, i.e. neglecting terms proportional to $\rho_{n+1}$ and keeping contributions $\rho_{n-1}$ \cite{schmidt2019}. For $n=1$ this corresponds to neglecting $\rho_2$ and keeping $\rho_0$ and $\rho_1$. This leads to
\begin{align}
\rho_n &\approx - \frac{1}{n^2 D_\text{rot}} \frac{\partial}{\partial x}  v_\text{f}  \frac{\rho_{n-1}}{2}, \label{eq:rhocutn}\\
\rho_1 &\approx - \frac{1}{ D_\text{rot}} \frac{\partial}{\partial x}  v_\text{f}  \left( \rho_{0} -\frac{\rho_1}{4} \right), \label{eq:rhocut1}
\end{align}
and thus only couples the expressions for $\rho_n$ with the lower order density coefficient $\rho_{n-1}$ for $n>1$. Hence one can, for a given density coefficient $\rho_n$, directly determine the next coefficient $\rho_{n+1}$ and then  iteratively calculate all higher order density coefficients $\rho_m$ for $m>n$.
 As a starting point of this simplified recursion relation, the orientationally averaged density $\rho_0$ is set to a hyperbolic tangent  
\begin{align}
\rho_0 = \frac{\rho_\text{l} + \rho_\text{g}}{2} + \frac{\rho_\text{l} - \rho_\text{g}}{2} \tanh(x/\lambda) \label{eq:rhotanh}
\end{align}
where  $\lambda$ indicates the width of the interface. The form of the density profile \eqref{eq:rhotanh} is a commonly used approximation in the ABP literature \cite{utrecht2018,rene2016,paliwal2017,vdW} and fits the simulation results very well.
Combining both assumptions, the form of $\rho_0$ \eqref{eq:rhotanh} and the truncation of the recursion relation, allows to determine the forward speed and iteratively all density coefficients. Hence the complete density distribution \eqref{eq:FTrho} and current distribution \eqref{eq:FTJx}, \eqref{eq:FTJy} using both \eqref{eq:Jxn} and \eqref{eq:Jx1} or both \eqref{eq:Jyn} and \eqref{eq:Jy1} can be calculated. The so derived expression for $v_\text{f}$ and the Fourier coefficients of $\rho$ and $\textbf{J}$ can then be compared to simulation results, see Ref. \cite{schmidt2019}.

We first consider the results for the gradient expression \eqref{eq:vf} of the forward speed $v_\text{f}$. 
It is also possible to express this quantity via the correlator \cite{schmidt2019}
\begin{align}
v_\text{f} = \frac{\int \textbf{J} \cdot \boldsymbol{\omega} \; \mathrm{d} \varphi}{\int \rho \; \mathrm{d} \varphi}, \label{eq:corrvf}
\end{align}
which can be proven by insertion of the current and the density Fourier decomposition, evaluation of the $\varphi$-integral and  using Eqs. \eqref{eq:Jx1} and \eqref{eq:Jy1}.
Thus $v_\text{f}(x)$ is the speed corresponding to the forward current,
\begin{align}
J_\text{f} = \rho_0 v_\text{f} = \frac{1}{2\pi} \int\limits_{0}^{2\pi} \textbf{J} \cdot \boldsymbol{\omega} \; \mathrm{d} \varphi, \label{eq:Jf}
\end{align}
which is the orientationally averaged projection of the one-body current $\textbf{J}$ on the particle ``forward'' direction $\boldsymbol{\omega}$, as the terminology already indicates. Since the orientationally integrated  density $2 \pi \rho_0 = \int \rho \mathrm{d} \varphi$ and the forward speed $v_\text{f}$ \eqref{eq:corrvf} are independent of the angle $\varphi$, the forward current $J_\text{f}(x)$ only depends on the spatial coordinate $x$. Recall that the current $\textbf{J}$ is in general not parallel to the orientation $\boldsymbol{\omega}$. 
The correlator expression \eqref{eq:corrvf} is useful to calculate the forward speed in the Brownian dynamic simulation from the measured current and density. (Alternatively and equivalently this speed can be calculated directly from a many-body expression \cite{krinninger2019}.) 

The forward velocity can be calculated explicitly from  Eq. \eqref{eq:vf} using the hyperbolic tangent profile \eqref{eq:rhotanh}. 
The calculation of the bulk densities as well as the determination of the value $\lambda$ is presented below in Sec. \ref{chap:phase}.
All remaining parameters were chosen via comparison to simulation data \cite{schmidt2019}.
The magnitude of $v_\text{f}$ changes smoothly from a high plateau value in the dilute state to a low value in the dense phase. This implies rapid particle movement in the gaseous phase and slow behaviour in the liquid state, which will be an important feature below in Sec. \ref{chap:phase} when addressing phase coexistence.

In the limit of a homogeneous bulk system the theoretical expression \eqref{eq:vf} reduces to the bulk forward speed
\begin{align}
v_\text{b} = v_\text{f}(\rho_0 \rightarrow \rho_\text{b}) = s \left( 1 - \frac{\rho_\text{b}}{\rho_\text{jam}} \right), \label{eq:vb}
\end{align}
decreasing linear in bulk density $\rho_\text{b}$. This can be seen by observing that gradient terms vanish for constant density $\rho = \rho_\text{b} = \text{const}$. This linear relationship is well-known \cite{solon2015,stenhammar2013,fily2012}. 
As maybe expected, this simplified, linear expression still describes with good accuracy both plateau values and those regions which asymptotically decay into bulk for large absolute values of $x/\sigma$.
In Appendix \ref{chap:bulk} the remaining bulk quantities of the system were determined starting from the bulk forward speed $v_\text{b}$.

Inserting the expression for the forward speed \eqref{eq:vf} and the 
hyperbolic tangent profile for the rotationally averaged density \eqref{eq:rhotanh} into the reduced recursion formula \eqref{eq:rhocutn} and \eqref{eq:rhocut1} allows to calculate the density coefficients iteratively.

The term $\rho_1$, which is proportional to the polarization (the $y$-component of the polarization vanishes due to symmetry), is peaked at the interface. Hence the swimmers at the interface tend to point towards the dense phase for the considered case of pure repulsive particles. This is reasonable since swimmers that are oriented towards the dilute phase have less sterical hindrance from other particles and their self-propulsion will make them ''escape`` into the gaseous phase. In contrast particles pointing towards the liquid phase impede colloids in the dense region from escaping into the dilute phase, which is a mechanism that is sometimes called self-trapping \cite{tailleur2009, bialke2015}. 
The second density Fourier coefficient $\rho_2$ can be interpreted as a nematic order parameter and shows a single oscillation in the interfacial region \cite{schmidt2019}. Thus the particle axis is parallel to the interface on the gas side and perpendicular to the interface on the liquid side.
Both effects decay towards the respective (isotropic) bulk. 

Note that we used truncated version of Eqs. \eqref{eq:Jxn}, \eqref{eq:Jx1} and \eqref{eq:Jyn}, \eqref{eq:Jy1} to calculate the current to avoid artefacts and to be consistent with the truncated recursion relation for the density. 
These truncated expressions \eqref{eq:JxnA}-\eqref{eq:Jy1AA} were derived together with the relation of the $y$-coefficients of the current $J_n^y$ to the density components $\rho_n$ \eqref{eq:JynA}, \eqref{eq:Jy1A} in Appendix \ref{chap:y}. 

We find that the magnitude of $\textbf{J}_1$ decreases from a high value in the gaseous to a lower value in the liquid phase. Its $y$-component has an additional peak at the interface. 
Hence there occurs a forward flux perpendicular to the density gradient at the interface of the system.

\subsection{Derivation of drag and superadiabatic pressure}
\label{chap:Fsup}

Having established physically correct and quantitatively reasonable solutions for the density and the current profiles, we proceed to determine the superadiabatic force densities that act in order to stabilize the interface. Recall that the superadiabatic force densities that contribute to the force density balance \eqref{eq:balance2}  have to be satisfied. As was the case for the forward speed, we define correlator expressions for the force densities $\textbf{F}_\text{sup,0}$, $\textbf{F}_\text{sup,1}$ and $\textbf{F}_\text{sup,2}$. The correlator expressions are all structurally similar, in the sense that they are proportional to orientationally integrated projections of $\textbf{F}_\text{int}$ on unit vectors. 
The spherical drag correlator \cite{schmidt2019} is
\begin{align}
\textbf{F}_\text{sup,0} = \frac{\textbf{J}}{2 \pi J_\text{f}} \int\limits_{0}^{2\pi} \textbf{F}_\text{int} \cdot \boldsymbol{\omega} \; \mathrm{d}  \varphi, \label{eq:corrFsup0}
\end{align}
and here $\boldsymbol{\omega}$ is the unit vector on which $\textbf{F}_\text{int}(x,\varphi)$ was projected. The expression for $\textbf{F}_\text{sup,0}$ is, as is the kinematic functional \eqref{eq:Fsup0}, proportional to the flux $\textbf{J}(x,\varphi)$. This implies that the direction of the force density is parallel to $\textbf{J}$. Furthermore $\textbf{F}_\text{sup,0}(x,\varphi)$ has no direct dependence on $\boldsymbol{\omega}(\varphi)$ and the indirect dependence on the particle orientation is completely contained in $\textbf{J}$,
since the forward current $J_\text{f}(x)$ and the integral in Eq. \eqref{eq:corrFsup0} are both independent of orientation due to integration in $\varphi$. However the integral depends on the $x$-position due to the dependence of the internal force density $\textbf{F}_\text{int}(x,\varphi)$ on position.

Using theoretical results, e.g. densities from the truncated recursion relation, one can determine the spherical drag force density from the kinematic functional  \eqref{eq:Fsup0} or from the correlator expression \eqref{eq:corrFsup0}.

 The $x$-component of $\textbf{F}_\text{sup,0}$ is directed against the particle flow. The $y$-component of $\textbf{F}_\text{sup,0}$ also acts against the current. The amplitude of both $x$- and $y$-component increases with increasing density. Both, the amplitude and the direction against the motion, show physically expected properties for a spherical drag force density.

The interfacial drag correlator $\textbf{F}_\text{sup,1}(x,\varphi)$ is given as
\begin{align}
\textbf{F}_\text{sup,1} = \frac{\boldsymbol{\omega}^*}{2 \pi} \int\limits_{0}^{2\pi} \left( \textbf{F}_\text{int} - \textbf{F}_\text{sup,0} \right) \cdot \boldsymbol{\omega}^* \; \mathrm{d}  \varphi, \label{eq:corrFsup1}
\end{align}
where the prefactor $\boldsymbol{\omega}^*(\varphi)$ generates the entire dependence on the particle orientation in $\textbf{F}_\text{sup,1}$, since the integral is performed with respect to $\varphi$. Note that before projection on the direction $\boldsymbol{\omega}^*$ and averaging with respect to the orientation, $\textbf{F}_\text{sup,0}(x,\varphi)$ is subtracted from the total internal force density $\textbf{F}_\text{int}(x,\varphi)$. Hence the non-spherical force density $\textbf{F}_\text{sup,1}$ constitutes a correction to the spherical drag contribution $\textbf{F}_\text{sup,0}$. 

An explicit result for the interfacial drag can be determined using Eq. \eqref{eq:Fsup1} or via insertion of $\textbf{F}_\text{int}$ \eqref{eq:Fint2} in Eq. \eqref{eq:corrFsup1} and neglecting the orientational dependence of the density in the ideal gas term.
We find $\textbf{F}_\text{sup,1}$ is an interfacial drag contribution, as it clearly is nonzero only in the interfacial region. The maximal amplitude is much smaller than the spherical drag $\textbf{F}_\text{sup,0}$, which confirms the status of $\textbf{F}_\text{sup,1}$ as a non-spherical correction. 
 
As already mentioned the superadiabatic force densities are kinematic expressions and it would be favourable to express the approximation \eqref{eq:Fsup1} for $\textbf{F}_\text{sup,1}$ without the usage of the swim contribution. Therefore we use Eq. \eqref{eq:vf} for the forward speed $v_\text{f}$ to replace the swim speed $s$ in \eqref{eq:Fsup1} and rewrite the relation as
\begin{align}
\textbf{F}_\text{sup,1} = - s \gamma \frac{\rho_1}{4} \boldsymbol{\omega}^* = \frac{\gamma v_\text{f} \rho_1}{4} \frac{1 + \xi (\nabla \rho_0)^2 \rho_0/\rho_\text{jam}}{1 - \rho_0/\rho_\text{jam}} \boldsymbol{\omega}^*. 
\end{align}

Similarly one can replace the dependence of $s$ in $\textbf{F}_\text{sup,2}$ \eqref{eq:Fsup2x},\eqref{eq:Fsub2y} (recall that this force density was determined to satisfy the force density balance) as 
\begin{align}
\textbf{F}_\text{sup,2} = - \gamma s \frac{\rho_1}{2} \; \hat{\textbf{e}}_x= - \frac{ \gamma v_\text{f} \rho_1}{2} \frac{1 + \xi (\nabla \rho_0)^2 \rho_0/\rho_\text{jam}}{1 - \rho_0/\rho_\text{jam}} \; \hat{\textbf{e}}_x.
\end{align}
Since this force density only contains a nonzero $x$-component and is independent of the $y$-coordinate due to symmetry, it may be written as the negative gradient of a scalar pressure $\Pi_2(x)$,
\begin{align}
\textbf{F}_\text{sup,2} = - \nabla \Pi_2. \label{eq:gradient}
\end{align}
In general the pressure is a tensor, of which the isotropic part is the scalar pressure. 
Note that our conventions are entirely consistent with continuum mechanics. 
The negative gradient of a pressure is identified as a force density and the negative gradient of a chemical potential is a force field. Rewriting the density coefficient $\rho_1$ with usage of the density recursion equation \eqref{eq:rho1} leads to the result 
\begin{align}
&\textbf{F}_\text{sup,2} = \frac{\gamma }{2 D_\text{rot}} \frac{\partial}{\partial x} \left[s v_\text{f}  \left(\rho_{0} -\frac{\rho_1}{4} + \frac{\rho_{2}}{2}\right) \right] \; \hat{\textbf{e}}_x \\
&= \frac{\gamma }{2 D_\text{rot}} \nabla \left[ v_\text{f}^2 \frac{1 + \xi (\nabla \rho_0)^2 \rho_0/\rho_\text{jam}}{1 - \rho_0/\rho_\text{jam}} \left(\rho_{0} -\frac{\rho_1}{4} + \frac{\rho_{2}}{2}\right) \right], \nonumber
\end{align}
where the spherical pressure is identified as
\begin{align}
\Pi_2 = - \frac{\gamma v_\text{f}^2 }{2 D_\text{rot}} \frac{1 + \xi (\nabla \rho_0)^2 \rho_0/\rho_\text{jam}}{1 - \rho_0/\rho_\text{jam}} \left(\rho_{0} -\frac{\rho_1}{4} + \frac{\rho_{2}}{2}\right). \label{eq:Pi2}
\end{align}
Analogously to the correlator expressions of the above considered superadiabatic drag force densities the correlator for $\textbf{F}_\text{sup,2}(x)$ is 
\begin{align}
\textbf{F}_\text{sup,2} = \frac{\hat{\textbf{e}}_x}{2 \pi} \int\limits_{0}^{2\pi} \textbf{F}_\text{int} \cdot \hat{\textbf{e}}_x \; \mathrm{d} \varphi. \label{eq:corrFsup2}
\end{align}
Note that insertion of the kinematic functionals and relation \eqref{eq:Fsup3} for the internal force density in all three correlators reproduces the correct equation for the respective superadiabatic force density, up to a thermal diffusion term, which is found to be numerically quite small. The deviation might be caused by the approximations for the superadiabatic force densities and could in principle be solved by introducing further superadiabatic terms (see Appendix \ref{chap:Fsup45}).

\subsection{Swim correlator expressions and swim pressure}
\label{chap:Fext}

All correlator expressions for the superadiabatic force densities \eqref{eq:corrFsup0}, \eqref{eq:corrFsup1} and \eqref{eq:corrFsup2} are proportional to the $\varphi$-integrated projection of the internal force density $\textbf{F}_\text{int}$ on another vector. For each of these equations there is a corresponding relation, where $\textbf{F}_\text{int}$ is replaced by the swim force density or self-propulsion term $\textbf{F}_\text{swim} = \gamma s \rho \boldsymbol{\omega}$. Although there are hardly any similarities between $\textbf{F}_\text{int}$ and $\textbf{F}_\text{swim}$, the force densities expressed by the correlators are quite similar expressions.

The force density corresponding to the spherical drag $\textbf{F}_\text{sup,0}$ \eqref{eq:corrFsup0} is the swim force density $\textbf{F}_\text{swim,0}$, given as
\begin{align}
\textbf{F}_\text{swim,0} = \frac{\textbf{J}}{2 \pi J_\text{f}} \int\limits_{0}^{2\pi} \textbf{F}_\text{swim} \cdot \boldsymbol{\omega} \; \mathrm{d}  \varphi. \label{eq:corrFext0}
\end{align}
Insertion of  $\textbf{F}_\text{swim}$ and explicit integration in $\varphi$ leads to
\begin{align}
\textbf{F}_\text{swim,0} = \frac{\gamma s}{2 \pi J_\text{f}} \textbf{J} \int\limits_{0}^{2\pi} \rho \; \mathrm{d} \varphi = \frac{\gamma s}{v_\text{f}} \; \textbf{J} = - \textbf{F}_\text{sup,0} + \gamma \textbf{J},
\end{align}
which shows the connection to $\textbf{F}_\text{sup,0}$.  

$\textbf{F}_\text{swim,1}$ corresponds to the interfacial drag contribution $\textbf{F}_\text{sup,1}$, see \eqref{eq:corrFsup1}, and is therefore defined as
\begin{align}
\textbf{F}_\text{swim,1} = \frac{\boldsymbol{\omega}^*}{2 \pi} \int\limits_{0}^{2\pi} \left( \textbf{F}_\text{swim} - \textbf{F}_\text{swim,0} \right) \cdot \boldsymbol{\omega}^* \; \mathrm{d}  \varphi. \label{eq:corrFext1}
\end{align}
Insertion of the the force densities and orientational integration demonstrates that $\textbf{F}_\text{swim,1}$ and $\textbf{F}_\text{sup,1}$ are identical up to a sign,  
\begin{align}
\textbf{F}_\text{swim,1} &= \frac{\boldsymbol{\omega}^*}{2 \pi} \gamma s \pi \rho_2 + \frac{\boldsymbol{\omega}^*}{2 \pi} \gamma s \pi \left(\frac{\rho_1}{2}-\rho_2 \right) = \frac{\gamma s \rho_1}{4}  \boldsymbol{\omega}^* \nonumber\\
&= - \textbf{F}_\text{sup,1}.
\end{align}
The contribution corresponding to $\textbf{F}_\text{sup,2}$ \eqref{eq:corrFsup2} is the second swim force density
\begin{align}
\textbf{F}_\text{swim,2}=  \frac{\hat{\textbf{e}}_x}{2 \pi} \int\limits_{0}^{2\pi} \textbf{F}_\text{swim} \cdot \hat{\textbf{e}}_x \; \mathrm{d} \varphi, \label{eq:corrFswim}
\end{align}
which is caused by the polarization $\pi \rho_1$ of the interface. Explicit integration of Eq. \eqref{eq:corrFswim} in $\varphi$ using the Fourier decomposition \eqref{eq:FTrho} leads to 
\begin{align}
\textbf{F}_\text{swim,2} =  \frac{\gamma s \rho_1}{2} \; \hat{\textbf{e}}_x = - \textbf{F}_\text{sup,2}, \label{eq:Fswp}
\end{align}
which directly implies $P_\text{swim}=-\Pi_2$ up to an irrelevant constant, such that
\begin{align}
\textbf{F}_\text{swim,2}  = - \nabla P_\text{swim}. \label{eq:Fswim}
\end{align}
As an aside, note that all introduced swim correlators $\textbf{F}_{\text{swim,}i}$ can be combined to the full expression for $\textbf{F}_\text{swim}$; hence $\textbf{F}_\text{swim}=\textbf{F}_\text{swim,0}+\textbf{F}_\text{swim,1}+\textbf{F}_\text{swim,2}$ holds. In case of the superadiabatic correlators there is no similar relation due to the thermal diffusion term $- k_\text{B} T \nabla \rho$. Neglecting of this ideal diffusion contribution in the correlator expressions results in $\textbf{F}_\text{sup,0}+\textbf{F}_\text{sup,1}+\textbf{F}_\text{sup,2} =\gamma \textbf{J} - \gamma s \rho \boldsymbol{\omega} =\textbf{F}_\text{int} - k_\text{B} T \nabla \rho$ as expected from Eq. \eqref{eq:Fint2}.

It is ''natural`` for the superadiabatic pressure caused by internal particle interactions to be independent of the swim speed $s$. Similarly the swim pressure which originates from self-propulsion should not contain the forward speed $v_\text{f}$. Using the relations \eqref{eq:Fswp}, \eqref{eq:Fswim} and \eqref{eq:Pi2} and replacing $v_\text{f}$ via Eq. \eqref{eq:vf} leads to the swim pressure
\begin{align}
P_\text{swim} = \frac{\gamma s^2}{2 D_\text{rot}} \frac{1 - \rho_0/\rho_\text{jam}}{1 + \xi (\rho_0)^2 \rho_0/\rho_\text{jam}} \left(\rho_{0} -\frac{\rho_1}{4} + \frac{\rho_{2}}{2}\right).
\end{align}
In order to construct a mathematically simple expression one could of course insert Eq. \eqref{eq:vf} only once and obtain the formula
\begin{align}
P_\text{swim} =-\Pi_2= \frac{\gamma s v_\text{f}}{2 D_\text{rot}} \left(\rho_{0} -\frac{\rho_1}{4} + \frac{\rho_{2}}{2}\right). \label{eq:Pswim}
\end{align}
In the limit of a bulk state this Eq. \eqref{eq:Pswim} simplifies to the previously obtained swim pressure 
\begin{align}
P_{\text{swim,}b}= \frac{\gamma s v_{b} \rho_b}{ 2 D_\text{rot}}.
\end{align} 
This expression is equivalent to the result respectively obtained in \cite{takatori2014,solon2015,rene2016,solon2018,yang2014,gompper2015}. 
The swim pressure is a central object in a variety of approaches  \cite{solon2015,takatori2014,gompper2015, speckjack, fily2018,gompper2019}.
For a detailed description of active bulk properties see Appendix \ref{chap:bulk}. 

The comparison between our swim pressure and the swim pressure from literature holds more generally than only in bulk. The origin of the term $\rho_1$ in Eq. \eqref{eq:Pswim} is the nonspherical drag contribution, which vanishes in the orientation-averaged case considered in \cite{takatori2014,solon2015,rene2016,solon2018,yang2014,gompper2015}. Therefore the contribution $\rho_1$ to the swim pressure could not be described in the above references. Furthermore the nematic contributions to the one-body density $\rho_2$ were neglected. 

To calculate the pressures via Eqs. \eqref{eq:Pi2} and \eqref{eq:Pswim}, the second density coefficient $\rho_2$ was neglected. This was done for consistency with the truncated recursion relation for the density coefficients.
In general $P_\text{swim}$ has a large contribution in the gas where the purely repulsive particles can swim more or less unimpeded. The swim pressure decreases with increasing density and a stronger effect of interparticle interactions from this high value in the gas to a low amplitude in the liquid phase. The behaviour of $\Pi_2$ is analogous but with opposite sign.

Summarizing, the corresponding force densities oppose and cancel each other,
as $P_\text{swim}$ \eqref{eq:Pswim} and $\Pi_2$ \eqref{eq:Pi2} are identical expressions (up to a minus sign). Thus the superadiabatic pressure $\Pi_2$ is balanced with the swim pressure $P_\text{swim}$, due to the relationship 
\begin{align}
P_\text{swim}+\Pi_2 =0,\label{eq:Pi}
\end{align}
so these partial pressures do not contribute to the total pressure.
The identity \eqref{eq:Pi} might be surprising at first glance. However, it justifies a posteriori the chosen splitting \eqref{eq:Fsup} of the superadiabatic force density $\textbf{F}_\text{sup}$.
To generate stable phase coexistence all pressure terms have to sum up to a constant value. Furthermore the sum of all chemical potential contributions has to add up to a constant. 
Due to Eq. \eqref{eq:Pi} we conclude that neither the swim pressure $P_\text{swim}$ nor the corresponding swim chemical potential $\mu_\text{swim}$ do contribute to the phase coexistence condition. This is in striking contrast to the findings of current literature \cite{takatori2014, solon2015, utrecht2018}, where $P_\text{swim}$ or $\mu_\text{swim}$ are claimed to induce MIPS.

\subsection{Construction of the phase diagram}
\label{chap:phase}
So far we have developed analytical expressions for the one-body density and the current distribution which solve approximately the force density balance \eqref{eq:balance} of the active system. In principle the expressions can be generalized using the full recursion relation and then satisfy the force density equation exactly. However there is as yet no condition that would determine the coexisting densities or the width of the interface, as is required for fully specifying the hyperbolic tangent form \eqref{eq:rhotanh}. 
To resolve this situation we use Eq. \eqref{eq:Fsup3} which was split off and not studied in detail so far. For simplicity, the force density balance \eqref{eq:Fsup3} is rewritten as a force balance and hence both its sides are divided by the one-body density field $\rho$, which yields
\begin{align}
\textbf{f}_\text{sup,3} - k_\text{B} T \nabla \ln \rho +\textbf{f}_\text{ad}= 0. \label{eq:force}
\end{align}
Here $\textbf{f}_\text{ad} = \textbf{F}_\text{ad}/\rho$ and $\textbf{f}_\text{sup,3} = \textbf{F}_\text{sup,3}/\rho$ denote the adiabatic and third superadiabatic force fields, respectively, and the thermal diffusion term was rewritten as $-k_\text{B} T (\nabla \rho)/\rho = - k_\text{B} T \nabla \ln \rho$. 
We approximate the ideal diffusive contribution as being independent of orientation and thus replace density $\rho$ via the orientationally averaged density $\rho_0$, so $- k_\text{B} T \nabla \ln \rho \approx - k_\text{B} T \nabla \ln \rho_0$ \cite{schmidt2019}. Therefore Eq. \eqref{eq:force} becomes
\begin{align}
\textbf{f}_\text{sup,3} - k_\text{B} T \nabla \ln \rho_0 +\textbf{f}_\text{ad}= 0. \label{eq:forceA}
\end{align}
Since the ideal contribution is negligibly small (its magnitude is about one order smaller than that of other terms \cite{schmidt2019}) and 
the gradient of $\rho$ is mainly influenced by the first Fourier coefficient $\rho_0$, this is a good approximation. 
In the reduced force balance \eqref{eq:forceA} the dependence on orientation of the superadiabatic force fields simplifies, as we will see below.
However the approximation can be in principle avoided and the balance \eqref{eq:forceA} directly obtained when additional superadiabatic force densities are taken into account (cf. Appendix \ref{chap:Fsup45}).

The approximated ideal chemical potential corresponding to the thermal diffusion term in \eqref{eq:forceA} is 
\begin{align}
\mu_\text{id} =  k_\text{B} T \ln \rho_0, \label{eq:muid}
\end{align}
up to an irrelevant constant.

Recall that the adiabatic force density contribution is determined via the density functional equation \eqref{eq:Fad}, or equivalently, expressed as a force field,
\begin{align}
\textbf{f}_\text{ad} = - \nabla \frac{\delta F_\text{exc}[\rho_0]}{\delta \rho_0}.
\end{align}
In the considered adiabatic reference system of spherical swimmers, the excess free energy functional $\textbf{F}_\text{exc}[\rho_0]$ is independent of orientation. Thus the adiabatic force field $\textbf{f}_\text{ad}$ is a gradient expression independent of $\varphi$. The force field can be expressed as 
\begin{align}
\textbf{f}_\text{ad} = - \nabla \mu_\text{ad},
\end{align}
where $\mu_\text{ad} = \delta F_\text{exc}[\rho_0]/\delta \rho_0$ denotes the adiabatic excess chemical potential. Since we base our local density approximation on scaled particle theory, $\mu_\text{ad}(x)$ is given as \cite{hansen}
\begin{align}
\mu_\text{ad} = k_\text{B} T \left[ - \ln (1-\eta') + \eta' \frac{3-2 \eta'}{(1-\eta')^2} \right], \label{eq:muad}
\end{align}
where $\eta'(x)=c \eta$ is a rescaled packing fraction $\eta = \rho_0/\rho_\text{jam}$ to approximately model the soft interparticle interaction potential and the possibility of the spherical swimmers to penetrate each other to some extent. The rescaling with factor $c\le 1$ is necessary, because the scaled particle theory assumes hard particles and allows to model various repulsive interparticle interactions.
Alternatively, $\eta'= \rho_0/\rho_\infty$ can be interpreted as a packing fraction with higher jamming density $\rho_\infty = \rho_\text{jam}/ c > \rho_\text{jam}$.

Since both the adiabatic force and the thermal diffusive contribution are gradient expressions and because of the chosen approximation independent of orientation, the superadiabatic force $\textbf{f}_\text{sup,3}$ has to be a gradient field and independent of $\varphi$ too, in order to satisfy the force balance \eqref{eq:forceA}. Hence $\textbf{f}_\text{sup,3}$ can be written in the form
\begin{align}
\textbf{f}_\text{sup,3} = - \nabla \nu_3, \label{eq:gradient2}
\end{align}
where $\nu_3(x)$ is a superadiabatic chemical potential, which is referred to as quiet life chemical potential \cite{schmidt2019}. 
The corresponding kinematic functional is chosen with a structure that is similar to that of the superadiabatic pressure $\Pi_2$ (cf.\ the quadratic dependence on speed and the linear dependence on density in the first term of $\nu_3$). 
Furthermore the quiet life potential has to balance the adiabatic chemical potential $\mu_\text{ad}$ (up to the small ideal contribution $k_\text{B}T \ln \rho_0$). 
We have chosen a simple and plausible form of $\nu_3$ and postulate this approximative relation to be \cite{schmidt2019}
\begin{align}
\nu_3 = \frac{\gamma}{2 D_{\text{rot}}} \Bigg[ &e_1 v^2_{\text{loc}} \frac{\rho_0}{\rho_{\text{jam}}} \nonumber \\
&+ \frac{e_2}{\rho^2_\text{jam}} \nabla \cdot \frac{ v^2_{\text{loc}}}{{\left( 1-\rho_0/\rho_{\text{jam}}\right)^2}} \nabla \rho_0 \Bigg], \label{eq:nu3}
\end{align}
where $e_1$ and $e_2$ are dimensionless constants. The local speed $v_{\text{loc}}$ is defined as 
\begin{align}
v_{\text{loc}} = v_{\text{f}} \left[ 1+ \xi (\nabla \rho_0)^2 \rho_0/\rho_{\text{jam}} \right]. \label{eq:vloc}
\end{align}
Keeping in mind that $\textbf{f}_\text{sup,3}$ is a kinematic functional  and therefore should be intrinsically independent of swim speed $s$, $v_\text{loc}(x)$ \eqref{eq:vloc} can be rewritten using Eq. \eqref{eq:vf} in the simple form  
\begin{align}
v_{\text{loc}} = s \left( 1- \frac{\rho_0}{\rho_\text{jam}} \right). 
\end{align}
Thus Eq. \eqref{eq:nu3} can be expressed as
\begin{align}
\nu_3 = \frac{s^2 \gamma}{2 D_{\text{rot}}} \left[ e_1 \left(1- \frac{\rho_0}{\rho_\text{jam}}\right)^2 \frac{\rho_0}{\rho_{\text{jam}}} + \frac{e_2}{\rho^2_\text{jam}} \nabla^2 \rho_0 \right]. \label{eq:v3}
\end{align}
Considering the structure of \eqref{eq:nu3} and \eqref{eq:v3} in more detail, $\nu_3$  consists of two contributions. The first term, a bulk contribution, will be relevant to determine the coexistence densities and has the same density and velocity dependence as has $\Pi_2$. 
The factor $e_1$ scales the magnitude of this term. 
The remaining second term in \eqref{eq:nu3} is an interfacial contribution, which vanishes in the bulk due to the absence of a gradient of $\rho_0$. It can be used to determine the width of the interface as is shown below. 
The magnitude of this interfacial contribution is scaled by $e_2$.
Although it would be desirable to derive the values $e_1$ and $e_2$ from first principles, these were obtained in Ref. \cite{schmidt2019} by comparison to simulation data.

The value of the combined chemical potential $\mu_\text{id} + \mu_\text{ad}$ increases from a negative value, generated by the ideal contribution, in the gas to a large positive value, primarily generated by the adiabatic contribution, in the liquid. As expected this chemical potential leads to a force towards the dilute phase and supports mixing of both phases.

The quiet life potential $\nu_3$ is high in the gas and low, approximately zero, in the liquid, since the corresponding superadiabatic force opposes both the adiabatic and ideal contribution. Therefore the term $\nu_3$ would induce phase separation. The name \textit{quiet life} was inspired from the observation that particles ''prefer`` the slow liquid phase (although there is more interparticle repulsion). Considering Eq. \eqref{eq:nu3} this effect is caused by the quadratic dependence on speed. The speed is low in the liquid and high in the gas. Therefore slow regions with small particle velocity are favoured, which we call the quiet life effect.

All three chemical potentials should add up to a constant value $\nu_3 + \mu_\text{ad} + \mu_\text{id}=\text{const}$, if the force balance \eqref{eq:forceA} is satisfied. 
Within our approximations this is indeed the case to a very satisfactory degree, see Fig. 5 in Ref. \cite{schmidt2019}. 

We are now in a position to introduce the total chemical potential $\mu_\text{tot}$, 
 given as the sum of all chemical potentials of the system,   
\begin{align}
\mu_\text{tot} = \mu_\text{id} + \mu_\text{ad}+\nu_3, \label{eq:mutot}
\end{align}
and the total pressure, 
\begin{align}
P_\text{tot} = P_\text{id} + P_\text{ad} + \Pi_3, \label{eq:ptot}
\end{align}
which is hence composed of the ideal pressure $P_\text{id}$, the adiabatic pressure $P_\text{ad}$ and the quiet life pressure $\Pi_3$.

For stable phase coexistence to occur both the total chemical potential $\mu_\text{tot}$and the total pressure $P_\text{tot}$ have to be constant.
These conditions, when applied to the two bulk phases, allow to determine coexistence densities by requiring 
\begin{align}
\mu_\text{tot}(\rho_g) = \mu_\text{tot}(\rho_l), \label{eq:tot1}\\
P_\text{tot}(\rho_g) = P_\text{tot}(\rho_l), \label{eq:tot2}
\end{align} 
and solving for $\rho_g$ and $\rho_l$. Recall the definition of the coexistence densities $\rho_g$ and $\rho_l$ below Eq. \eqref{eq:gibbs}. 
Since the interfacial contributions do not contribute in Eqs. \eqref{eq:tot1} and \eqref{eq:tot2}, we only need to consider bulk terms, evaluated at a bulk density $\rho_\text{b}=\rho_g$ or $\rho_l$. 
In case of $\nu_3(\rho_\text{b})$ the second contribution in \eqref{eq:nu3} vanishes and the local speed simplifies to the bulk speed $v_\text{b}$ \eqref{eq:vb}, so 
\begin{align}
\nu_3(\rho_\text{b}) = \frac{e_1 \gamma}{2 D_{\text{rot}}}  v^2_\text{b} \frac{\rho_\text{b}}{\rho_{\text{jam}}}. \label{eq:nu3b}
\end{align}
Insertion of all contributions into Eq. \eqref{eq:mutot} leads to 
\begin{align}
\mu_\text{tot}(\rho_\text{b}) = \; &k_\text{B}T \ln \eta'_\text{b} - k_\text{B} T  \ln (1-\eta'_\text{b}) \\ 
&+ c k_\text{B} T \frac{\rho_\text{b}}{\rho_\text{jam}} \frac{3-2 \eta'_b}{(1-\eta'_\text{b})^2}+ \frac{e_1 \gamma}{D_{\text{rot}}}  v^2_\text{b} \frac{\rho_\text{b}}{\rho_{\text{jam}}}. \nonumber 
\end{align}
where we have used $\eta'_\text{b}= c \rho_\text{b} / \rho_\text{jam}$.
There are no further contributions from the corresponding chemical potentials of $\Pi_2$ and $P_\text{swim}$, 
since both terms cancel each other up to an irrelevant offset. 

The total pressure can be determined from the total chemical potential by integrating in $\rho_\text{b}$ the Gibbs-Duhem equation 
\begin{align}
\frac{\partial P_\text{tot}}{\partial \rho_\text{b}} = \rho_\text{b} \frac{\partial \mu_\text{tot}}{\partial \rho_\text{b}}, \label{eq:thermo}
\end{align}
which holds by construction. 
Again the swim pressure $P_\text{swim}$ does not contribute to the total pressure since it is cancelled by the superadiabatic pressure $\Pi_2$. 
The remaining contributions to the total pressure \eqref{eq:ptot} are derived by separately solving the Gibbs-Duhem equation \eqref{eq:thermo} for the ideal \eqref{eq:muidb}, adiabatic \eqref{eq:muadb} or superadiabatic \eqref{eq:nu3b} bulk chemical potential.
One finds \cite{schmidt2019}
\begin{align}
P_\text{id}(\rho_\text{b}) &= k_\text{B} T \rho_\text{b}, \label{eq:Piid} \\
P_\text{ad}(\rho_\text{b}) &= k_\text{B} T \rho_\text{b} \left[ \frac{1}{(1-\eta'_\text{b})^2} - 1 \right], \label{eq:Piad}\\
\Pi_3(\rho_\text{b}) &= \frac{e_1 \gamma}{4 D_\text{rot}} v_\text{b}^2 \frac{\rho_\text{b}^2}{\rho_\text{jam}} \left[ 1 + \frac{\rho_\text{b} \left( 3 \rho_\text{b} - 4 \rho_\text{jam} \right)}{6(\rho_\text{b} - \rho_\text{jam})^2} \right]. \label{eq:Pi3}
\end{align}

Having expressions for the total pressure and the total chemical potential Eqs. \eqref{eq:tot1} and \eqref{eq:tot2} can be solved numerically for the bulk densities $\rho_g$ and $\rho_l$ at coexistence. The results for the densities as a function of the Péclet number $\text{Pe}$ constitute the theoretical phase diagram for MIPS.
 The dimensionless Péclet number relates the strength of active swimming against that of diffusion, 
\begin{align} 
\text{Pe} = \frac{s \gamma \sigma}{k_\text{B} T} = \frac{3 s}{\sigma D_\text{rot}}.
\end{align}
Therefore, $\text{Pe}$ measures the scaled activity of the particles. 
The coexistence densities for different values of $\text{Pe}$ form the binodal. One finds that the density of the active gas has even for rather high values of $\text{Pe}$ a significant value well removed from zero density. This is in contrast to typical equilibrium gas-liquid phase diagrams where $\rho_g$ approaches very dilute densities with increasing inverse temperature.

The spinodal is set via the densities that satisfy the condition $\partial \mu_\text{tot}(\rho_\text{b})/\partial \rho_\text{b} = 0$
and shows surprisingly high values of density for the gaseous branch in comparison to the behaviour in equilibrium. These properties for the binoal and spinodal have also been observed in simulations \cite{utrecht2018,stenhammar2013}.

The critical point determines the onset of phase separation and it is given when both the first and second derivative of the total chemical potential vanish. Hence $\partial \mu_\text{tot}(\rho_\text{b})/\partial \rho_\text{b} = 0$ as for the spinodal and additionally $\partial^2 \mu_\text{tot}(\rho_\text{b})/\partial \rho_\text{b}^2 = 0$. These conditions lead to a fourth order polynomial which is derived explicitly in Appendix \ref{chap:crit}. The relevant zero of this polynomial can be obtained numerically, see Fig. 2 in \cite{schmidt2019} for the result.
For a simulation study of the critical behaviour of active Brownian particles see \cite{speck2018,dietrich2020}.

Note that in our derivation of the phase diagram no interfacial contributions were necessary. Recall that the total pressure and  the total chemical potential have to be constant not only in bulk but also at the interface for a stable interface to exist and therefore a stable phase separation to occur \cite{schmidt2019}. 
Since the conditions for the total pressure and chemical potential are related via Eq. \eqref{eq:thermo} a perfectly constant chemical potential would imply a constant pressure. In bulk this stability condition is satisfied by the choice of the coexistence densities. At the interface the fulfilment of this relation can be controlled via the amplitude of the interfacial term in $\nu_3$. Its magnitude is proportional to $e_2/\lambda^2$ due to the second derivative of $\rho_0$ with respect to $x$. Therefore the amplitude can be fitted such that $\mu_\text{tot}$ is as constant as possible and from this amplitude can be derived the interfacial width $\lambda$ for a fixed value of $e_2$.

\section{Conclusions and outlook} 
\label{chap:conclusions} 

We have given a full account of the theoretical description developed in Refs. \cite{schmidt2019,postension} for phase separating active Brownian particles. The underlying formal framework of Krinninger et al.\ \cite{krinninger2016, krinninger2019} was complemented by concrete and physically plausible kinematic force approximations.
 It was shown that the positional and orientationally resolved force density balance splits into two distinct contributions. 
The first contribution \eqref{eq:balance2} (using \eqref{eq:Fint2}) contains the friction term, self-propulsion, spherical and non-spherical drag force densities as well as the superadiabatic pressure gradient. This contribution was used together with the continuity equation to derive a recursion relation for the one-body density Fourier coefficients. 
The second contribution \eqref{eq:Fsup3} to the force density balance contains the adiabatic term, the ideal gas contribution and the quiet life term. We have shown that it is the second contribution that determines phase coexistence. The bulk densities at coexistence were calculated and the interfacial width follows from matching the magnitude of the quiet life force with the magnitude of the adiabatic and ideal forces.

The active gas-liquid binodal is determined starting from a chemical potential and a pressure using the Maxwell construction similar to \cite{brady2015,utrecht2018}. In Refs. \cite{solon2015,utrecht2018} it is a common conclusion for high particle activity that the phase coexistence cannot be determined from Maxwell construction. The authors of \cite{utrecht2018,solon2018} argue that interfacial contributions have to be taken into account due to the anisotropy of the system \cite{utrecht2018}. In contrast to this conclusion we find that the phase diagram obtained by the equal area construction matches the simulation data to a very satisfying degree. 
Due to Eq. \eqref{eq:Pi}, the superadiabatic and the swim pressure, arising from the interfacial particle polarization, cancel each other pointwise across the interface. Hence these two pressure contributions and similarly the corrsponding chemical potential contributions do not contribute to the stability conditions \cite{schmidt2019}.
We assumed that the total pressure and chemical potentials are constant across the interface. This is indeed satisfied to a very good degree, see Fig. 5 in Ref. \cite{schmidt2019} for a graphical representation.
Furthermore, the polarization is found to be a state function in this system \cite{sumrule}, so bulk quantities alone determine the total polarization at the interface.
Therefore we conclude that a coupling from the interface to the bulk due to the occurring particle polarization is not necessary for the description. This is the same situation as in equilibrium. Furthermore our approximate expression for the polarization, in the form \eqref{eq:rho1} or \eqref{eq:rhocut1}, satisfies the exact polarisation sum rule of Ref. \cite{sumrule}. The normal and tangential components of the pressure tensor vary differently across the interface and the integral over the difference constitutes the interfacial tension, which is found to be positive, see Ref. \cite{postension}.

The stable nonequilibrium phase coexistence originates from repulsive forces: the quiet life force compresses the liquid and the ideal and adiabatic force expands the liquid and therefore compress the gaseous phase. It would be interesting to investigate further the relationship with the theory of Farage et~al. \cite{farage2015} and whether the obtained repulsion can be reinterpretated 
as an effective attractive interparticle force. 

The swim pressure which arises due to the particle polarization at the interface is proportional to the orientation-averaged density, as well as to the swim speed and to the density dependent particle speed. In Refs. \cite{brady2015,solon2015} the pressure is claimed to be important for the formation of phase separation. We have shown that the swim pressure does not contribute to coexistence, since this term cancels exactly with the (second) superadiabatic pressure. The superadiabatic pressure arises from the corresponding superadiabatic force density, which was introduced to satisfy the first contribution of the force density balance.

To summarize, the considered system can now be investigated with standard Maxwell phase coexistence tools. 
Despite its simple form the developed theory captures all important observed physical phenomena such as the phase separation and the surprisingly high density of the gas phase. Considering the necessary approximations, as e.g. a local density approximation and mean-field theory, and the very small number of five fit parameters (three for the bulk and two more for the interfacial description) the agreement with Brownian dynamics simulations is very satisfying, in particular for the phase diagram \cite{schmidt2019}.

Our treatment works on the level of both fully orientation- and position-resolved one-body fields. Hence the influence of external torques on the particles lies within the realm of our treatment forming possible future work.
It would be worthwhile to extend the theory from two to three spatial dimensions \cite{stenhammar2014,gompper2014}, see Refs. \cite{wilding,geissler} for recent studies, in particular of freezing. In three dimensions the particle orientation varies on a unit sphere (not a unit circle). Therefore the Fourier expansion of the density and the current \eqref{eq:FTrho}-\eqref{eq:FTJy} becomes an expansion in spherical harmonics which makes calculations more involved.
It might be interesting to extend the framework to nonlocal contributions and to include memory \cite{treffenstaedt2020,jahreis2020}, possible aided by the nonequilibrium Ornstein-Zernike relations \cite{noz1,noz2} and the Noether theorem for statistical physics \cite{noether}.   
Furthermore, it would be fruitful to investigate sedimentation \cite{sedimentation}, active-passive mixtures \cite{filion1,valeriani,marenduzzo2015,gompper2016,klapp12020}, dipolar active Brownian particles \cite{klapp2016,klapp2020}, and curved interfacial shapes \cite{auschra2021}.
It would be highly interesting to extend the theory to Lennard-Jones interactions, where very different phenomena (from MIPS in purely repulsive particles) were reported \cite{paliwal2017,filion2015, filion1,filion2}. This can be useful to examine the applicability and generality of the developed concepts. 

Future applications of our framework could well be motivated by the study of confined systems, such a freezing of water inside of pore structures \cite{findenegg,jaehnert,schreiber} and capillary condensation \cite{thommes,roecken}, that Findenegg and his coworkers pioneered. 
 Given the experimental control of colloidal particles that e.g. nudging with light \cite{cichos,cichos1,cichos2,cichos3,cichos4} and topological transport by magnetic fields \cite{fischer2016,fischer2020} allows, thinking into such directions seems both reasonable and exciting.

\textit{Acknowledgements.} This paper is dedicated to the memory of Gerhard Findenegg. We would like to thank Philip Krinninger, Jeroen Rodenburg, Siddarth Paliwal, Marjolein Dijkstra, Ren\'e van Roij, Ren\'e Wittmann, Hartmut Löwen, Thomas Speck, Sven Auschra, Klaus Kroy, Victor Holubec, Nicola Söker, Frank Cichos, Gerhard Gompper, Grzegorz Szamel, Julien Tailleur, Martin Oettel, Bob Evans, Joseph Brader, and Thomas Fischer for inspiring comments and useful discussions. We are indebted to Umberto Marconi for pointing out typographical errors in two equations. The work has been supported by the German Research Foundation (DFG) under project number 447925252.

\appendix

\section{Active Brownian bulk fluids}
\label{chap:bulk}

The density distribution in bulk is homogeneous, thus $\rho(x, \varphi) = \rho_\text{b} = \text{const}$ is independent of position $x$ and orientation $\boldsymbol{\omega}$. The speed is constant and it is assumed to decrease linear in density
\begin{align}
v_\text{b} = s \left( 1 - \frac{\rho_b}{\rho_\text{jam}} \right) \label{eq:8}
\end{align}
in accordance with simulations and literature \cite{menzel2014,solon2015,stenhammar2013,fily2012}.
In the low density limit this behaviour can be derived by considering the particle collisions, 
cf.\ Ref. \cite{stenhammar2013}.
The absolute value of the current is $J_\text{b} = \rho_\text{b} v_\text{b}$. The direction of the current is along to the  particle orientation $\boldsymbol{\omega}$ since the (averaged) interparticle interactions are the same from all directions and hence have no influence on the one-body current. The vectorial current is given as
\begin{align}
\textbf{J}_\text{b} = \rho_\text{b} v_\text{b} \boldsymbol{\omega} = s \rho_\text{b} \left( 1 - \frac{\rho_b}{\rho_\text{jam}} \right)\boldsymbol{\omega}. \label{eq:7}
\end{align}
The force density balance \eqref{eq:balance} simplifies in bulk to 
\begin{align}
\gamma \textbf{J}_\text{b} = \textbf{F}_\text{sup,b} + s\gamma \rho_\text{b} \boldsymbol{\omega}, \label{eq:6}
\end{align}
where the ideal diffusive term and the adiabatic force density vanish since both contain a spatial derivative.
The superadiabatic bulk force density follows from inserting \eqref{eq:7} in Eq. \eqref{eq:6} as 
\begin{align}
\textbf{F}_\text{sup,b} = - s \gamma \frac{\rho_\text{b}^2}{\rho_\text{jam}} \boldsymbol{\omega} = - \gamma \frac{\rho_\text{b}}{\rho_\text{jam} - \rho_\text{b}} \textbf{J}_\text{b}. \label{eq:9}
\end{align}
In the second step this force density was rewritten to emphasize the connection to the spherical drag contribution $\textbf{F}_\text{sup,0}$ \eqref{eq:Fsup0}. Since all other superadiabatic force densities (determined in Sec. \ref{chap:mips}) vanish in bulk, it is logical that only $\textbf{F}_\text{sup,0}$ contributes to $\textbf{F}_\text{sup,b}$. Furthermore the correlator expressions are exact in bulk, because the ideal diffusion contribution vanishes due to the constant density distribution $\rho_\text{b}$. Note that as $\textbf{F}_\text{sup,b}$ \eqref{eq:Fsup0} simplifies in bulk to \eqref{eq:9} the forward speed \eqref{eq:vf} and the current \eqref{eq:FTJx}, \eqref{eq:FTJy} also reduce to the above derived relations \eqref{eq:8}, \eqref{eq:7} in the bulk limit.

Although the adiabatic force density $\textbf{F}_\text{ad}$ and the superadiabatic contributions $\textbf{F}_{\text{sup,}n}$ for $n=1,2,3$ vanish in bulk fluids, the corresponding pressures and chemical potentials 
are in general different from zero. 
The superadiabatic pressure $\Pi_2$ \eqref{eq:Pi2} reduces to 
\begin{align}
\Pi_2(\rho_\text{b}) &= - \frac{\gamma v_\text{b}^2}{2 D_\text{rot}} \frac{1}{1 - \rho_\text{b}/\rho_\text{jam}} \rho_\text{b}  \nonumber \\
&= - \frac{\gamma s^2}{2 D_\text{rot}} \rho_\text{b} \left(1 - \rho_\text{b}/\rho_\text{jam}\right)
\end{align}
in the limit of a bulk fluid. 
For the swim contributions applies $\Pi_2(\rho_\text{b}) = - P_\text{swim}(\rho_\text{b})$ as before and hence
\begin{align}
P_\text{swim}(\rho_\text{b}) = \frac{\gamma s^2}{2 D_\text{rot}} \rho_\text{b} \left(1 - \rho_\text{b}/\rho_\text{jam}\right) = \frac{\gamma}{2 D_\text{rot}} s v_\text{b} \rho_\text{b}.
\end{align}
The determined bulk swim pressure is linear in density $\rho_\text{b}$, linear in the swim speed $s$ and linear in the density dependent bulk speed $v_\text{b}(\rho_\text{b})$ and thus equal to the pressure obtained in the literature \cite{takatori2014,solon2015,rene2016,solon2018,yang2014,gompper2015}.
The corresponding chemical potential can be determined using the Gibbs-Duhem equation $\partial \Pi_2 /\partial \rho_\text{b} = \rho_\text{b} \partial \nu_2 /\partial \rho_\text{b}$. The result is
\begin{align}
\nu_2(\rho_\text{b}) = - \mu_\text{swim}(\rho_\text{b}) = \frac{\gamma s^2}{2 D_\text{rot}} \left( \frac{2 \rho_\text{b}}{\rho_\text{jam}} - \ln \rho_\text{b} \right).
\end{align}

In the derivation of the phase coexistence conditions (see Sec. \ref{chap:phase}) we have already obtained the bulk contributions for the ideal, the adiabatic and the quiet life term. For completeness the relations are rewritten in the following. 

The ideal contribution contains the ideal pressure \eqref{eq:Piid} and the ideal chemical potential \eqref{eq:muid},
\begin{align}
P_\text{id}(\rho_\text{b}) &= k_\text{B} T \rho_\text{b}, \\
\mu_\text{id}(\rho_\text{b}) &=  k_\text{B}T \ln \rho_\text{b}. \label{eq:muidb}
\end{align}
The adiabatic terms, the adiabatic pressure \eqref{eq:Piad} and adiabatic chemical potential \eqref{eq:muad}, are in scaled particle theory given as
\begin{align}
P_\text{ad}(\rho_\text{b}) = k_\text{B} T \rho_\text{b} &\left[ \frac{1}{(1-c \rho_\text{b}/\rho_\text{jam})^2} - 1 \right], \\
\mu_\text{ad}(\rho_\text{b}) = k_\text{B} T \bigg[ - &\ln \left(1-c \frac{\rho_\text{b}}{\rho_\text{jam}}\right) \nonumber\\
+ &c \frac{\rho_\text{b}}{\rho_\text{jam}} \frac{3-2 c \; \rho_\text{b}/\rho_\text{jam}}{(1-c \; \rho_\text{b}/\rho_\text{jam})^2} \bigg]. \label{eq:muadb}
\end{align}
The quiet life pressure \eqref{eq:Pi3} and the quiet life chemical potential \eqref{eq:nu3b} were found as 
\begin{align}
&\Pi_3(\rho_\text{b}) = \nonumber\\
&=\frac{e_1 \gamma s^2}{4 D_\text{rot}} \frac{\rho_\text{b}^2}{\rho_\text{jam}} \left( 1 - \frac{\rho_\text{b}}{\rho_\text{jam}} \right)^2 \left[ 1 + \frac{\rho_\text{b} \left( 3 \rho_\text{b} - 4 \rho_\text{jam} \right)}{6(\rho_\text{b} - \rho_\text{jam})^2} \right] \nonumber\\
&=\frac{e_1 \gamma}{4 D_\text{rot}} v_\text{b}^2 \frac{\rho_\text{b}^2}{\rho_\text{jam}} \left[ 1 + \frac{\rho_\text{b} \left( 3 \rho_\text{b} - 4 \rho_\text{jam} \right)}{6(\rho_\text{b} - \rho_\text{jam})^2} \right], \\
&\nu_3(\rho_\text{b}) = \frac{e_1 \gamma s^2}{D_{\text{rot}}}  \frac{\rho_\text{b}}{\rho_{\text{jam}}} \left( 1 - \frac{\rho_\text{b}}{\rho_\text{jam}} \right)^2 = \frac{e_1 \gamma}{D_{\text{rot}}}  v^2_\text{b} \frac{\rho_\text{b}}{\rho_{\text{jam}}}.
\end{align}

\section{Derivation of the recursion relation for the $y$-component of the current}
\label{chap:y}
The derivation of expressions for the $y$-component current Fourier coefficients is analogous to the calculation for $x$-components (cf. Sec.~\ref{chap:recursion}). The starting point is the $y$-component of the force density balance \eqref{eq:balance}
\begin{align}
\gamma J^y = \gamma s \rho \sin \varphi + F^y_\text{int}.
\end{align}
Expressing the internal force density via Eq. \eqref{eq:Fint2} and inserting the assumed, approximative relations for the drag force densities $F^y_\text{sup,0}$ \eqref{eq:Fsup0} and $F^y_\text{sup,1}$ \eqref{eq:Fsup1} yields
\begin{align}
\gamma J^y = \; &\gamma s \rho \sin \varphi - \gamma \frac{\rho_0}{\rho_\text{jam} - \rho_0} \left[ 1 + \xi (\nabla \rho_0)^2 \right] J^y \nonumber \\
&- \gamma s \frac{\rho_1}{4} \sin\varphi + F_\text{sup,2}^y.
\end{align}
Performing a the Fourier decomposition of the density and the current using Eqs. \eqref{eq:FTrho} and \eqref{eq:FTJy} leads to
\begin{align}
&\gamma \sum\limits_{n=1}^\infty J_n^y \sin(n \varphi) \label{eq:test}\\
&= \; \gamma s \sum\limits_{n=0}^\infty \rho_n \cos(n \varphi) \sin \varphi - \gamma s \frac{\rho_1}{4} \sin\varphi + F_\text{sup,2}^y  \nonumber \\
&{\color{white} = } - \gamma \frac{\rho_0}{\rho_\text{jam} - \rho_0} \left[ 1 + \xi (\nabla \rho_0)^2 \right] \sum\limits_{n=1}^\infty J_n^y \sin(n \varphi). \nonumber
\end{align}
The self-propulsion contribution (first term on the right hand side) can be expressed as
\begin{align}
&\gamma s \sum\limits_{n=0}^\infty \rho_n \cos(n \varphi) \sin \varphi \label{eq:sin}\\
&= \sum\limits_{n=0}^\infty \frac{\gamma s}{2} \rho_n \Biggl[  \sin((n+1) \varphi) - \sin((n-1) \varphi) \Biggr] \nonumber\\
&= \frac{\gamma s}{2} \left[ \sum\limits_{n=1}^\infty \rho_{n-1} \sin( n \varphi) - \sum\limits_{n=-1}^\infty \rho_{n+1} \sin( n \varphi) \right] \nonumber \\
&= \frac{\gamma s}{2} \left[\frac{\rho_0}{2} \sin \varphi +  \sum\limits_{n=1}^\infty (\rho_{n-1} - \rho_{n+1}) \sin(n \varphi)  \right], \nonumber
\end{align}
using the trigonometric relation $2 \cos(n \varphi) \sin \varphi =  \sin((n+1) \varphi) - \sin((n-1) \varphi)$.
Insertion of the relation \eqref{eq:sin} and rewriting Eq. \eqref{eq:test} in orders of $\sin(n \varphi)$ gives
\begin{align}
 F_\text{sup,2}^y =  &- \gamma s \left( \frac{\rho_0}{2} + \frac{\rho_1}{4} \right) \sin \varphi  \label{eq:test2}\\
 &+ \gamma \sum\limits_{n=1}^\infty \left[ J_n^y \frac{s}{v_\text{f}} - \frac{s}{2} \left( \rho_{n-1} - \rho_{n+1} \right) \right] \sin(n \varphi).\nonumber
\end{align}
Since the terms $\sin (n \varphi)$ with different values of $n$ are independent of each other, it is required for the corresponding prefactors to vanish in order to satisfy the equation. For $n>1$ this requirement corresponds to 
\begin{align}
J^y_n  = \frac{v_\text{f}}{2}  \left(\rho_{n-1} - \rho_{n+1}\right), \label{eq:JynA}
\end{align}
obtained from setting the square brackets in Eq. \eqref{eq:test2} equal to zero. In case of $n=1$ an additional term has to be taken into account. As in the $x$-component it is caused by the first superadiabatic force density $F^y_\text{sup,1}$ and a sum boundary term of the self-propulsion. This leads to
\begin{align}
J^y_1 = v_\text{f} \left(\rho_{0} +\frac{\rho_{1}}{4} - \frac{\rho_{2}}{2} \right). \label{eq:Jy1A}
\end{align}
Except for $F^y_\text{sup,2}$ there are no further contributions independent of $\boldsymbol{\omega}$, so we conclude
\begin{align}
F_\text{sup,2}^y = 0. \label{eq:Fsub2y}
\end{align}
The determined relations for the $n$th Fourier coefficient of the current $J^y$ \eqref{eq:JynA}, \eqref{eq:Jy1A} are structurally similar to those for the $x$-component $J^x$ \eqref{eq:Jxn}, \eqref{eq:Jxn}. They differ only in sign of the highest density distribution $\rho_{n+1}$ or for $n=1$ in sign of $\rho_1$ and $\rho_2$.

Using the same truncation for the current components as for the density in Sec. \ref{chap:rho}, i.e. neglecting the highest order in  density $\rho_{n+1}$, yields
\begin{align}
J^x_n  &\approx v_\text{f} \; \frac{ \rho_{n-1}}{2},\label{eq:JxnA}\\
J^x_1 &\approx v_\text{f} \left(\rho_{0} -\frac{\rho_{1}}{4} \right),\\
J^y_n  &\approx v_\text{f} \; \frac{\rho_{n-1}}{2},\\
J^y_1 &\approx v_\text{f} \left(\rho_{0} +\frac{\rho_{1}}{4} \right). \label{eq:Jy1AA}
\end{align}
One can clearly see that within this approximation $J^x_n = J^y_n$ for $n>1$.

\section{Superadiabatic force densities $\textbf{F}_\text{sup,4}$ and $\textbf{F}_\text{sup,5}$}
\label{chap:Fsup45}

The correlators introduced in Sec. \ref{chap:Fsup} agree with the approximated kinematic functionals only up to the thermal diffusion term. This restriction can be avoided if one introduces two additional superadiabatic force densities $\textbf{F}_\text{sup,4}$ and $\textbf{F}_\text{sup,5}$ as 
\begin{align}
\textbf{F}_\text{sup,4} &= k_\text{B} T \nabla \rho, \label{eq:Fsup4}\\
\textbf{F}_\text{sup,5} &= k_\text{B} T \frac{\rho}{\rho_0} \nabla \rho_0. \label{eq:Fsup5}
\end{align}
Note that correlators and kinematic functionals of these force densities are as yet unknown. The terms are defined via the contribution that they cancel analogous to $\textbf{F}_\text{sup,3}$, which was introduced by \eqref{eq:Fsup3} and later on specified as \eqref{eq:gradient2} and \eqref{eq:nu3}.

In the following it is shown that the force density balance still holds and that the correlators agree with the kinematic functionals. The internal force density $\textbf{F}_\text{int} = \textbf{F}_\text{ad} +\textbf{F}_\text{sup}$ is extended by the new contributions to
\begin{align}
\textbf{F}_\text{int} = &\; \textbf{F}_\text{sup,0}+\textbf{F}_\text{sup,1}+\textbf{F}_\text{sup,2}+\textbf{F}_\text{sup,3}\nonumber\\
&+\textbf{F}_\text{sup,4}+\textbf{F}_\text{sup,5}+\textbf{F}_\text{ad}
\end{align}
Inserting this relation in the force density balance \eqref{eq:balance} leads to
\begin{align}
\gamma \textbf{J} = & \; \textbf{F}_\text{int} - k_\text{B} T \nabla \rho + s \gamma \rho \boldsymbol{\omega} \nonumber\\
= & \; \textbf{F}_\text{sup,0}+\textbf{F}_\text{sup,1}+\textbf{F}_\text{sup,2}+\textbf{F}_\text{sup,3} \nonumber\\
&+\textbf{F}_\text{sup,4}+\textbf{F}_\text{sup,5}+\textbf{F}_\text{ad} - k_\text{B} T \nabla \rho + s \gamma \rho \boldsymbol{\omega} \nonumber\\
= & \;\textbf{F}_\text{sup,0}+\textbf{F}_\text{sup,1}+\textbf{F}_\text{sup,2}+\textbf{F}_\text{sup,3} \nonumber \\
&+\textbf{F}_\text{sup,5}+\textbf{F}_\text{ad} + s \gamma \rho \boldsymbol{\omega}, \label{eq:5}
\end{align}
in which Eq. \eqref{eq:Fsup4} was applied in the last step. Defining the third superadiabatic force as in Eq. \eqref{eq:forceA} gives, rewritten as a force density balance and using \eqref{eq:Fsup5}, the relation
\begin{align} 
0 &= \textbf{F}_\text{sup,3} - k_\text{B} T \frac{\rho}{\rho_0} \nabla \rho_0 +\textbf{F}_\text{ad} \nonumber \\
&=\textbf{F}_\text{sup,3} +\textbf{F}_\text{sup,5} +\textbf{F}_\text{ad}. \label{eq:4}
\end{align}
Insertion of Eq. \eqref{eq:4} in the force density balance \eqref{eq:5} yields
\begin{align}
\gamma \textbf{J} &= \textbf{F}_\text{sup,0}+\textbf{F}_\text{sup,1}+\textbf{F}_\text{sup,2}+\textbf{F}_\text{sup,4} - k_\text{B} T \nabla \rho + s \gamma \rho \boldsymbol{\omega} \nonumber \\
&= \textbf{F}_\text{sup,0}+\textbf{F}_\text{sup,1}+\textbf{F}_\text{sup,2} + s \gamma \rho \boldsymbol{\omega}.
\end{align}
This balance is identical to Eq. \eqref{eq:balance2} with \eqref{eq:Fint2} as required. 
That the approximative theoretical relations satisfy the correlator expressions, the internal force density $\textbf{F}_\text{int}$ in the correlators \eqref{eq:corrFsup0}, \eqref{eq:corrFsup1} and \eqref{eq:corrFsup2} has to be replaced by 
\begin{align}
\textbf{F}'_\text{int} &= \textbf{F}_\text{int} - \textbf{F}_\text{sup,4} \nonumber \\
&= \textbf{F}_\text{sup,0}+\textbf{F}_\text{sup,1}+\textbf{F}_\text{sup,2}+\textbf{F}_\text{sup,3}+\textbf{F}_\text{sup,5}+\textbf{F}_\text{ad} \nonumber \\
&= \textbf{F}_\text{sup,0}+\textbf{F}_\text{sup,1}+\textbf{F}_\text{sup,2}.
\end{align}
In this case the new $\textbf{F}'_\text{int}$ is equal to the previous internal force density \eqref{eq:Fint2} when the thermal diffusion contribution is neglected.

\section{Critical point}
\label{chap:crit}

At the critical point the first and second derivative of $\mu_\text{tot}$ \eqref{eq:mutot} with respect to $\rho_0$ vanish, thus 
\begin{align}
\frac{\partial \mu_\text{tot}}{\partial \rho_\text{b}} &= \frac{\partial \mu_\text{id}}{\partial \rho_\text{b}} + \frac{\partial \mu_\text{ad}}{\partial \rho_\text{b}} + \frac{\partial \nu_3}{\partial \rho_\text{b}} = 0, \label{eq:sum1}\\
\frac{\partial^2 \mu_\text{tot}}{\partial \rho_\text{b}^2} &= \frac{\partial^2 \mu_\text{id}}{\partial \rho_\text{b}^2} + \frac{\partial^2 \mu_\text{ad}}{\partial \rho_\text{b}^2} + \frac{\partial^2 \nu_3}{\partial \rho_\text{b}^2} = 0. \label{eq:sum2}
\end{align}
First all occurring derivatives were evaluated using the relation $\partial \eta_\text{b}/\partial \rho_\text{b} = 1/\rho_\text{jam}$, which yields
\begin{align}
\frac{\partial (\mu_\text{id} + \mu_\text{ad})}{\partial \rho_\text{b}} &= \frac{k_\text{B} T}{\rho_\text{jam}} \frac{1 + c \eta_\text{b}}{\eta_\text{b} (1 - c \eta_\text{b})^3}, \label{eq:mu1}\\
\frac{\partial \nu_3}{\partial \rho_\text{b}}&= \frac{k_\text{B} T}{\rho_\text{jam}} \frac{e_1 \text{Pe}^2}{6} (1 - \eta_\text{b})(1 - 3 \eta_\text{b}), \label{eq:nu1}\\
\frac{\partial^2 (\mu_\text{id}+\mu_\text{ad})}{\partial \rho_\text{b}^2}&= \frac{k_\text{B} T}{\rho_\text{jam}^2} \frac{3 c^2 \eta_0^2 + 4 c \eta_\text{b} -1}{\eta_\text{b}^2 (1 - c \eta_\text{b})^4}, \label{eq:mu2}\\
\frac{\partial^2 \nu_3}{\partial \rho_\text{b}^2}&= \frac{k_\text{B} T}{\rho_\text{jam}^2} \frac{e_1 \text{Pe}^2}{6} (6 \eta_\text{b} - 4). \label{eq:nu2}
\end{align}
Equation \eqref{eq:sum1} can be rewritten inserting Eq. \eqref{eq:mu1} and \eqref{eq:nu1} as
\begin{align}
\frac{e_1 \text{Pe}^2}{6} \eta_0 (1 - c \eta_\text{b})^3 = - \frac{1 + c \eta_\text{b}}{(1- \eta_\text{b})(1 - 3\eta_\text{b})} \label{eq:1}
\end{align}
and Eq. \eqref{eq:sum2} gives 
\begin{align}
\frac{e_1 \text{Pe}^2}{6} \eta_0 (1 - c \eta_\text{b})^3 = - \frac{3 c^2 \eta_\text{b}^2 + 4 c \eta_\text{b} -1}{ \eta_\text{b} (1- c \eta_\text{b})(6\eta_\text{b} - 4)} \label{eq:2}
\end{align}
using \eqref{eq:mu2} and \eqref{eq:nu2}.
Since both relations were reordered such that they have the same contributions on left hand side, on can equate Eqs. \eqref{eq:1} and \eqref{eq:2}. From this follows
\begin{align}
&(1-c \eta_\text{b})(1+c\eta_\text{b})(6\eta_\text{b} - 4)\eta_\text{b} \nonumber\\
&= (1- \eta_\text{b})(1-3\eta_\text{b})(3 c^2 \eta_\text{b}^2 + 4 c \eta_\text{b} -1) \label{eq:3}
\end{align}
and hence after multiplying out the fourth order polynomial
\begin{align}
0= &- 15 c^2 \eta_\text{b}^4 + (16c^2-12c) \eta_\text{b}^3 + (-3c^2-16c+9)\eta_\text{b}^2 \nonumber\\
&+ (-4c-8)\eta_\text{b} + 1.
\end{align}
Due to the fundamental theorem of algebra this polynomial has four zeros. 
The desired packing fraction $\eta_\text{b,crit}$ can be identified via physical properties as $\eta_\text{b} > 0$ and comparison with the phase diagram. 
An expression for the critical Péclet number $\text{Pe}_\text{crit}$ can be determined from 
\begin{align}
\text{Pe}_\text{crit}^2 e_1 = - \frac{1+c \eta_\text{0,crit}}{\eta_\text{0,crit} (1-c \eta_\text{0,crit})^3 (1 - \eta_\text{0,crit}) (1 - 3 \eta_\text{0,crit})},
\end{align}
which was derived by insertion of the critical packing fraction $\eta_\text{b,crit}$ in Eq. \eqref{eq:3}.



\begin{thebibliography}{99}

\bibitem{schmidt2019} S. Hermann, P. Krinninger, D. de las Heras, and M. Schmidt, Phys. Rev. E \textbf{100}, 052604 (2019).

\bibitem{postension} S. Hermann, D. de las Heras, and M. Schmidt, Phys. Rev. Lett. \textbf{123}, 26802 (2019).

\bibitem{krinninger2019} P. Krinninger and M. Schmidt, J. Chem. Phys \textbf{150}, 074112 (2019).

\bibitem{krinninger2016} P. Krinninger, M. Schmidt, and J. M. Brader, Phys. Rev. Lett. \textbf{117}, 208003 (2016).

\bibitem{rev1} G. Gonnella, D. Marenduzzo, A. Suma, and A. Tiribocchi, C. R. Phys. \textbf{16}, 316 (2015).

\bibitem{rev2} T. Speck, Eur. Phys. J. Special Topics \textbf{225}, 2287 (2016).

\bibitem{rev3} S. C. Takatori and J. F. Brady, Curr. Opin. Colloid Interface Sci. \textbf{21}, 24 (2016).

\bibitem{rev4} G. Gompper et al., J. Phys.: Condens. Matter \textbf{32}, 193001 (2020).

\bibitem{rev5} M. C. Marchetti, J. F. Joanny, S. Ramaswamy, T. B. Liverpool, J. Prost, M. Rao, and R. A. Simha, Rev. Mod. Phys. \textbf{85}, 1143 (2013).

\bibitem{rev6} M. E. Cates and J. Tailleur, Annu. Rev. Condens. Matter Phys. \textbf{6}, 219 (2015).

\bibitem{rev7}  
A. Zöttl and H. Stark, J. Phys.: Condens. Matter \textbf{28}, 253001 (2016).


\bibitem{stark2014}
A. Zöttl and H. Stark, Phys. Rev. Lett. \textbf{112}, 118101 (2014).


\bibitem{gompper2018} 
M. Theers, E. Westphal, K. Qi, R. G. Winkler, and G. Gompper, Soft Matter \textbf{14}, 8590 (2018).


\bibitem{bialke2013} J. Bialké, H. L{\"o}wen, and T. Speck, EPL \textbf{103}, 30008 (2013).

\bibitem{menzel2014} T. Speck, J. Bialké, A. M. Menzel, and H. L{\"o}wen, Phys. Rev. Lett. \textbf{112}, 218304 (2014).

\bibitem{menzel2015} T. Speck, A. M. Menzel, J. Bialké, and H. L{\"o}wen, J. Chem. Phys., \textbf{142}, 224109 (2015).

\bibitem{takatori2014} S. C. Takatori, W. Yan, and J. F. Brady, Phys. Rev. Lett. \textbf{113}, 028103 (2014).

\bibitem{brady2015} S. C. Takatori and J. F. Brady, Phys. Rev. E \textbf{91}, 032117 (2015).

\bibitem{brady2020} S. A. Mallory, A. K. Omar, and J. F.  Brady, arXiv:2009.06092

\bibitem{stenhammar2014} J. Stenhammar, D. Marenduzzo, R. J. Allen, and M. E. Cates,  Soft Matter \textbf{10}, 1489 (2014).

\bibitem{solon2015}  A. P. Solon, J. Stenhammar, R. Wittkowski, M. Kardar, Y. Kafri, M. E. Cates, and J. Tailleur, Phys. Rev. Lett. \textbf{114}, 198301 (2015).

\bibitem{neta2020} P. D. Neta, M. Tasinkevych, M. M. Telo da Gama, and C. S. Dias, 
 Soft Matter {\bf 17}, 2468 (2021).

\bibitem{utrecht2018} S. Paliwal, J. Rodenburg, R. van Roij, and M. Dijkstra, New J. Phys. \textbf{20}, 015003 (2018).

\bibitem{farage2015} T. F. F. Farage, P. Krinninger, and J. M. Brader, Phys. Rev. E \textbf{91}, 042310 (2015).

\bibitem{fox} R. F. Fox, Phys. Rev. A \textbf{33}, 467 (1986); R. F. Fox, Phys. Rev. A \textbf{34}, 4525(R) (1986).

\bibitem{rene2016} R. Wittmann and J. M. Brader, EPL \textbf{114}, 68004 (2016).

\bibitem{negTension2015} J. Bialké, H. L{\"o}wen, and T. Speck, Phys. Rev. Lett. \textbf{115}, 098301 (2015).

\bibitem{speck2016} T. Speck, EPL \textbf{114}, 30006 (2016).

\bibitem{bradytension} A. K. Omar, Z.-G. Wang, and J. F. Brady, Phys. Rev. E \textbf{101}, 012604 (2020).

\bibitem{speck2020} T. Speck, Soft Matter \textbf{16}, 2652 (2020).

\bibitem{speck12020} T. Speck, Phys. Rev. E \textbf{103}, 012607 (2021).

\bibitem{solon2018} A. P. Solon, J. Stenhammar, M. E. Cates, Y. Kafri, and J. Tailleur, New J. Phys. \textbf{20}, 075001 (2018).

\bibitem{pft2013} M. Schmidt and J. M. Brader, J. Chem. Phys. \textbf{138}, 214101 (2013).

\bibitem{fortini2014} A. Fortini, D. de las Heras, J. Brader, and M. Schmidt, Phys. Rev. Lett. \textbf{113}, 167801 (2014).

\bibitem{sumrule} S. Hermann and M. Schmidt, Phys. Rev. Research \textbf{2}, 022003(R) (2020).

\bibitem{auschra2020}  S. Auschra, V. Holubec, N. A. Söker, F. Cichos, and K. Kroy, arXiv:2010.16234 .

\bibitem{soeker2020} N. A. Söker, S. Auschra, V. Holubec, K. Kroy, and F. Cichos, arXiv:2010.15106.

\bibitem{geigenfeind} T. Geigenfeind, D. de las Heras, and M. Schmidt, Comms. Phys. \textbf{3}, 23 (2020).

\bibitem{dzubiella2002} J. Dzubiella, G. P. Hoffmann, and H. Löwen, Phys. Rev. E \textbf{65}, 021402 (2002).

\bibitem{dzubiella2003} J. Chakrabarti, J. Dzubiella, and H. Löwen, Europhys. Lett. \textbf{61}, 415 (2003).

\bibitem{dzubiella2004} J. Chakrabarti, J. Dzubiella, and H. Löwen, Phys. Rev. E \textbf{70}, 012401 (2004).

\bibitem{treffenstaedt} L. L. Treffenstädt and M. Schmidt, Soft Matter \textbf{16}, 1518 (2020).

\bibitem{jahreis2020} N. Jahreis and M. Schmidt, Col. Pol. Sci. \textbf{298}, 895 (2020).

\bibitem{prl2020} D. de las Heras and M. Schmidt, Phys. Rev. Lett. \textbf{125}, 018001 (2020).

\bibitem{velocitygradient} D. de las Heras and M. Schmidt, Phys. Rev. Lett. \textbf{120}, 028001 (2018).

\bibitem{stuhlmueller2018} N. C. X. Stuhlm{\"u}ller, T. Eckert, D. de las Heras, and M. Schmidt, Phys. Rev. Lett. \textbf{121}, 098002 (2018). 

\bibitem{renner2019} D. de las Heras, J. Renner, and M. Schmidt, Phys. Rev. E \textbf{99}, 023306 (2019).

\bibitem{vitelli} B. C. van Zuiden, J. Paulose, W. T. M. Irvine, D. Bartolo, and V. Vitelli, Proc. Natl. Acad. Sci. U.S.A. \textbf{113}, 12919 (2016).

\bibitem{loewen} A. Härtel, R. Blaak, and H. Löwen, Phys. Rev. E \textbf{81}, 051703 (2010).

\bibitem{loewen2} M. Marechal, H. H. Goetzke, A. Härtel, and H. Löwen, J. Chem. Phys. \textbf{135}, 234510 (2011).

\bibitem{stenhammar2013} J. Stenhammar, A. Tiribocchi, R. J. Allen, D. Marenduzzo, and M. E. Cates, Phys. Rev. Lett. \textbf{111}, 145702 (2013).

\bibitem{sedimentation} S. Hermann and M. Schmidt, Soft Matter \textbf{140}, 1614 (2018).

\bibitem{philipDis} P. Krinninger, \textit{Effective Equilibrium, Power Functional, and Interface Structure for Phase-Separating Active Brownian Particles}, Ph.D. Thesis, Universit{\"a}t Bayreuth (2018).

\bibitem{evans1979} R. Evans, Adv. Phys. \textbf{28}, 143 (1979).

\bibitem{evans2016} R. Evans, M. Oettel, R. Roth, and G. Kahl, J. Phys.: Condens. Matter \textbf{28}, 240401 (2016).

\bibitem{pagona2018} P. Digregorio, D. Levis, A. Suma, L. F. Cugliandolo, G. Gonnella, and I. Pagonabarraga, Phys. Rev. Lett. \textbf{121}, 098003 (2018).

\bibitem{paliwal2017} S. Paliwal, V. Prymidis, L. Filion, and M. Dijkstra, J. Chem. Phys. \textbf{147}, 084902 (2017).

\bibitem{vdW} J. D. van der Waals, Z. Phys. Chem. \textbf{13}, 657  (1894); J. S. Rowlinson, J. Stat. Phys. \textbf{20}, 197 (1979) (translation).

\bibitem{fily2012} Y. Fily and M. C. Marchetti, Phys. Rev. Lett. \textbf{108}, 235702 (2012).

\bibitem{tailleur2009} J. Tailleur and M. E. Cates, Europhys. Lett. \textbf{86}, 60002 (2009).

\bibitem{bialke2015} J. Bialké, T. Speck, and H. Löwen, J. Non-Cryst. Solids \textbf{407}, 367 (2015).

\bibitem{yang2014} X. B. Yang, L. M. Manning, and M. C. Marchetti, Soft Matter \textbf{10}, 6477 (2014).

\bibitem{gompper2015}
R. G. Winkler, A. Wysocki and G. Gompper, Soft Matter \textbf{11}, 6680 (2015).


\bibitem{speckjack}
T. Speck and R. L. Jack, Phys. Rev. E \textbf{93}, 062605 (2016).


\bibitem{fily2018}
Y. Fily, Y. Kafri, A. P. Solon, J. Tailleur, and A. Turner, J. Phys. A: Math. Theor. \textbf{51}, 044003 (2018).

\bibitem{gompper2019}
S. Das, G. Gompper, and R. G. Winkler, Sci. Rep. \textbf{9}, 6608 (2019).

\bibitem{hansen} J.-P. Hansen and I. R. McDonald \textit{Theory of Simple Liquids} (Academic Press, Amsterdam, 2013).

\bibitem{speck2018} J. T. Siebert, F. Dittrich, F. Schmid, K. Binder, T. Speck, and P. Virnau, Phys. Rev. E \textbf{98}, 030601 (2018).

\bibitem{dietrich2020} F. Dittrich, T. Speck, P. Virnau, arxiv:2010.08387.

\bibitem{gompper2014}
A. Wysocki, R. G. Winkler, and G. Gompper, EPL \textbf{105}, 48004 (2014).

\bibitem{wilding}
F. Turci and N. B. Wilding, Phys. Rev. Lett. \textbf{126}, 038002 (2021).


\bibitem{geissler}
A. K. Omar, K. Klymko, T. GrandPre, and P. L. Geissler, arXiv:2012.09803.

\bibitem{treffenstaedt2020} L. L. Treffenstädt and M. Schmidt (unpublished).
%

\bibitem{noz1} J. M. Brader and M. Schmidt, J. Chem. Phys. \textbf{139}, 104108 (2013).

\bibitem{noz2} J. M. Brader and M. Schmidt, J. Chem. Phys. \textbf{140}, 034104 (2014).

\bibitem{noether} S. Hermann and M. Schmidt (submitted).

\bibitem{filion1} B. van der Meer, V. Prymidis, M. Dijkstra, and L. Filion, J. Chem. Phys. \textbf{152}, 144901 (2020). 

\bibitem{valeriani} D. R. Rodriguez, F. Alarcon, R. Martinez, J. Ramírez, and C. Valeriani, Soft Matter \textbf{16}, 1162 (2020).

\bibitem{marenduzzo2015} 
J. Stenhammar, R. Wittkowski, D. Marenduzzo, and M. E. Cates, Phys. Rev. Lett. \textbf{114}, 018301 (2015).


\bibitem{gompper2016}
A. Wysocki, R. G. Winkler, and G. Gompper, New J. Phys. \textbf{18}, 123030 (2016).


\bibitem{klapp12020}
R. C. Maloney, G.-J. Liao, S. H. L. Klapp, and C. K. Hall, Soft Matter \textbf{16}, 3779 (2020).


\bibitem{klapp2016}
S. H. L. Klapp, Curr. Opin. Colloid Interface Sci. \textbf{21}, 76 (2016).


\bibitem{klapp2020} 
G.-J. Liao, C. K. Hall, and S. H. L. Klapp, Soft Matter \textbf{16}, 2208 (2020).


\bibitem{auschra2021}
S.  Auschra and  V.  Holubec,  arXiv:2101.04548.


\bibitem{filion2015} V. Prymidis, H. Sielcken and L. Filion, Soft Matter \textbf{11}, 4158 (2015).

\bibitem{filion2} V. Prymidis, S. Paliwal, M. Dijkstra, and L. Filion, J. Chem. Phys. \textbf{145}, 124904 (2016). 

\bibitem{findenegg} G. H. Findenegg, S. J{\"a}hnert, D. Akcakayiran, and A. Schreiber, ChemPhysChem \textbf{9}, 2651 (2008).

\bibitem{jaehnert} S. J{\"a}hnert, V. Chavez, G. E. Schaumann, A. Schreiber, M. Sch{\"o}nhoff, and G. H. Findenegg, Phys. Chem. Chem. Phys. \textbf{10}, 6039 (2008).

\bibitem{schreiber} A. Schreiber, I. Ketelsen, and G. H. Findenegg, Phys. Chem. Chem. Phys. \textbf{3}, 1185 (2001).

\bibitem{thommes} M. Thommes and G. H. Findenegg, Langmuir \textbf{10}, 4270 (1994).

\bibitem{roecken} P. R{\"o}cken, A. Somoza, P. Tarazona, and G. Findenegg, J. Chem. Phys. \textbf{108}, 8689 (1998).

\bibitem{cichos} U. Khadka, V. Holubec, H. Yang, F. Cichos, Nat. Commun. \textbf{9}, 3864 (2018).

\bibitem{cichos1} M. Selmke, U. Khadka, A. P. Bregulla, F. Cichos, and H. Yang,  Phys. Chem. Chem. Phys. \textbf{20}, 10502 (2018).

\bibitem{cichos2} M. Selmke,U. Khadka, A. P. Bregulla, F. Cichos, and H. Yang,  Phys. Chem. Chem. Phys. \textbf{20}, 10521 (2018).

\bibitem{cichos3} A. P. Bregulla, H. Yang, and F. Cichos, ACS Nano \textbf{8}, 6542 (2014).

\bibitem{cichos4} B. Qian, D. Montiel, A. Bregulla, F. Cichos, and H. Yang, Chem. Sci. \textbf{4}, 1420 (2013).

\bibitem{fischer2016} J. Löhr, M. Lönne, A. Ernst, D. de las Heras, and T. M. Fischer, Nat. Commun. \textbf{7}, 11745 (2016).

\bibitem{fischer2020} M. Mirzaee-Kakhki, A. Ernst, D. de las Heras, M. Urbaniak, F. Stobiecki, J. Gördes, M. Reginka, A. Ehresmann, and T. M. Fischer, Nat. Commun. \textbf{11}, 4670 (2020).

\end{thebibliography}
\end{document}